\documentclass[journal,onecolumn]{IEEEtran}
\makeatletter
\usepackage{cite}
\usepackage{amsmath,amssymb,amsfonts,bm,ulem}
\usepackage{amsthm}
\usepackage{mathrsfs}
\usepackage{algorithmic}
\usepackage{graphicx}
\usepackage{textcomp}
\usepackage{epstopdf}
\usepackage{lipsum}
\makeatother

\ifCLASSOPTIONcompsoc
\usepackage[caption=false, font=normalsize, labelfont=sf, textfont=sf]{subfig}
\else
\usepackage[caption=false, font=footnotesize]{subfig}

\theoremstyle{remark}
\newtheorem{theorem}{Theorem}
\newtheorem{lemma}{Lemma}  
  
\newtheorem{definition}{Definition}

\newtheorem{proposition}{Proposition}

\begin{document}

\title{Active Anomaly Detection with Switching Cost}

\author{Fengfan~Qin,
		Da~Chen,
        Hui~Feng,~\IEEEmembership{Member,~IEEE},
        Qing~Zhao,~\IEEEmembership{Fellow,~IEEE},
        Tao~Yang,~\IEEEmembership{Member,~IEEE},
        and~Bo~Hu,~\IEEEmembership{Member,~IEEE}  
    
\thanks{F. Qin, D. Chen, H. Feng, T. Yang and B. Hu are with the Research Center of Smart Networks and Systems, School of Information Science and Engineering, 
Fudan University, Shanghai 200433, China. e-mail: (\{ffqin18, dachen16, hfeng, taoyang, bohu\}@fudan.edu.cn,). A priliminary result was presented in the ICASSP 2019-2019 IEEE International Conference on Acoustics, Speech and Signal Processing (ICASSP) \cite{ChenDa}. }
\thanks{Q. Zhao is with School of Electrical and Computer Engineering, Cornell University. e-mail:qz16@cornell.edu}
}


\maketitle

\begin{abstract}

The problem of detecting a single anomalous process among multiple independent processes is considered. Under a constraint on the number of processes that can be probed simultaneously, the decision maker  should decide which processes to probe at each time and when to terminate the probing. Compared with previous work considering only the observation costs, the switching costs of switchings across processes also need to be taken into account in many practical scenarios. The objective is an active inference strategy that minimizes the Bayesian risk taking into account of the sample complexity, switching cost, as well as detection errors. Based on the framework of sequential design of experiments, we propose a low-complexity, low-switching deterministic policy for two scenarios where the total switching cost is negligible and the total switching cost is comparable to the total observation cost. We show that the proposed algorithm is asymptotically optimal in the former scenario and is order optimal in the latter scenario. Simulation results demonstrate strong performance in the finite regime for both scenarios.
\end{abstract}

\begin{IEEEkeywords}
Active hypothesis testing, anomaly detection, switching cost.
\end{IEEEkeywords}

\IEEEpeerreviewmaketitle

\section{Introduction}

\IEEEPARstart{C}{onsider} the problem of detecting anomalous processes among multiple processes. At each time, only one process can be probed and a noisy observation is obtained from the process. The objective is to find an optimal inference strategy consisting of a selection rule indicating which process to probe at each time, a stopping rule on when to terminate the detection, and a decision rule on the final result. The performance measure is the Bayesian risk that takes into account various costs during procedure, such as the detection error, the sampling delay or switching cost, etc.

The anomaly detection problem studied in this paper relates to the sequential design of experiments problem first studied by Chernoff in 1959 \cite{cd_1} named as the Active Hypothesis Testing  (AHT) problem. Differ from the classic sequential hypothesis testing pioneered by Wald in \cite{Wald} where the observation action under each hypothesis is fixed, active hypothesis testing allows the decision maker to choose different experiments to be conducted at each time and infer the state of the probed process.  The test developed by Chernoff (referred to as Chernoff test) is a \emph{randomized} test that generates random actions based on historical observations. The actions, consequently the test statistics of log likelihood ratio, thus become independent and identically distributed over time, allowing asymptotic analysis of the stopping time on the test statistic.

Chernoff's work has been extended in various directions. Bessler \cite{Bessler} generalized Chernoff's work to general multiple hypothesis testing in 1960. Naghshvar and Javidi in  \cite{Naghshvar_2010, Naghshvar_2013, Naghshvar-information, Naghshvar-performance, Naghshvar-sequential} studied this sequential problem from the perspective of minimizing Bayesian risk considering the detection error and detection delay. In \cite{Nitinawarat-2012, Nitinawarat-2013, Nitinawarat-2015}, more general models which considered Markovian observations and non-uniform costs on actions were proposed. Recently, K. Cohen and Q. Zhao et al. studied AHT for anomaly detection in \cite{cohen_bayes, cohen_cost1, cohen_tree, cohen_cost2, cohen_hete, cohen_costnonline, AHT2020-KOBI}. In contrast to the random policies advocated in other works, they introduced a simple deterministic model which offers the same asymptotic optimality yet with significant performance gain in the finite regime and considerable reduction in implementation complexity. More recent studies and applications of general active hypothesis testing problem can be found in \cite{AHT2014, AHT2017-1, AHT2020-1, AHT2020-2}.

However, Chernoff test and subsequent studies do not consider switching cost. Incorporating switching cost into Bayesian risk is motivated by a number of applications. For example, in many searching tasks, relocating the searching agent (eg. rescuing ships or air-crafts) incurs considerable cost in terms of energy  or delay. Another example is medical diagnostics, where frequent and fast switching across drugs and medical procedures may carry high risk and side effects.

There are few studies considering the switching cost in AHT. Vaidhiyan developed a modified Chernoff test (referred to as Sluggish policy) in \cite{Sluggish} and introduced a switching parameter $\eta$ which determines the switching probability, which can be seen as an $\varepsilon$-greedy strategy. They claimed that Sluggish policy approaches the asymptotic performance of Chernoff test as $\eta \rightarrow 0$. But this policy results in a higher observation cost in the finite regime as demonstrated in their simulations.

In this paper, we propose a low-complexity deterministic test for the above active hypothesis testing problem with switching cost, referred to the Deterministic Bounded Switching (DBS) policy, which was first formulated and studied in our prior work \cite{ChenDa}. The proposed policy explicitly specifies the probing action at each time based on the history of observations, which integrates all parameters affecting the Bayesian risk: the number of processes  $M$, the single-switching cost $s$, the single-observation cost $c$, and the corresponding Kullback-Liebler (KL) divergences between the observation distributions of normal and anomalous processes. DBS policy partitions the problem space into two cases. In the first case, the process that are most likely to be abnormal will be probed. In the other case, DBS policy probes processes that are likely to be normal and eliminates them one by one to reduce the number of switchings. In addition to the scenario studied in \cite{ChenDa}  where the single-switching cost is of no greater order than the single-observation cost in the asymptotic regime of $c\rightarrow 0$, we further analyze the performance of the DBS policy in the scenario where the total switching cost is negligible and the total switching cost is comparable to the total observation cost. Then, we show that the DBS policy enjoys asymptotic optimality when the total switching cost is negligible to the observation cost, i.e., $s=o(-c\log c)$ as $c$ approaches $0$, and strong performance in the finite regime are demonstrated in simulation part of this paper. In addition, we also analyze the performance of the DBS policy when total switching cost is comparable to the total observation cost, i.e., $s=\Omega(-c\log c)$ as $c$ approches $0$, and draw the conclusion that the DBS policy is order optimal in this scenario.


In this paper, Sec. \ref{model} describes the system model for single anomaly detection problems. DBS policy is proposed to solve it, and its optimality is proved in Sec. \ref{DBS-optimal}. In Sec. \ref{DGF} and Sec. \ref{SPRT} we compare DBS with the deterministic DGF policy in \cite{cohen_bayes} and the Random Sequential Probability Ratio Test (R-SPRT)  policy and analyze the impact of switching cost. In Sec. \ref{simulation} we provide numerical examples to illustrate the performance of the proposed policy compared with other algorithms. Sec. \ref{conclusion} concludes the paper.

\section{SYSTEM MODEL AND PROBLEM FORMULATION}
\label{model}
\subsection{System Model}
Consider the problem of detecting an anomalous process among a finite number (denoted by $M$) of processes. Each process may be in a normal state or an abnormal state alternatively. Let $H_m$ denote the hypothesis that the process $m$ is in an abnormal state. Let $\pi_m$ be the priori probability that $H_m$ is true, where $\sum_{m=1}^{M}{\pi}_{m}=1$, and $0<{\pi}_{m}<1$ for all $m$. At each given time, only one process can be probed. When process $m$ is probed at time $n$, an observation $y_m(n)$ is obtained and is independent over time. If process $m$ is in a normal state, $y_m(n)$ follows distribution $f(y)$; if process $m$ is in an abnormal state, $y_m(n)$ follows distribution $g(y)$. We focus on the case where the distributions $f(y)$ and $g(y)$ are known.

Let $P_{m}$ be the probability measure under hypothesis ${H}_{m}$ and $E_{m}$ be the operator of expectation concerning the measure $P_{m}$. Let $\phi(n)\in {1,2,...,M}$ be a selection rule indicating which process is probed at time $n$. The vector of selection rules is denoted by $\bm{\phi}=(\phi(n),n\geq 1)$. Let $\tau$ be a stopping time (or a stopping rule), which is the time when the decision maker finishes the detecting procedure. Let $\tau_c$ and $\tau_s$ be the numbers of observations and switchings by the stopping time $\tau$, respectively, we have $\tau=\tau_c+\tau_s$. Let $\delta \in \{ 1,2,...,M \}$ be a decision rule, where $\delta=m$ if the decision maker declares that ${H}_{m}$ is true at time $\tau$.

An admissible policy for the sequential anomaly detection problem is denoted by
\begin{equation}
\Gamma=(\tau,\delta,\bm{\phi}).
\end{equation}

\subsection{Objective}
Let ${\emph P}_{e}(\Gamma)\triangleq \sum_{m=1}^{M}{\pi}_{m}{\alpha}_{m}(\Gamma)$ be the probability of error under policy $\Gamma$, where ${\alpha}_{m}(\Gamma) \triangleq P_{m}(\delta \neq m|\Gamma)$ is the probability of declaring $\delta \neq m$ when ${H}_{m}$ is true. Let $E(\tau | \Gamma )\triangleq \sum_{m=1}^{M}{\pi}_{m}E_{\emph m}(\tau|\Gamma)$ be the expected detection time under $\Gamma$, while $E(\tau_{c} | \Gamma )\triangleq \sum_{m=1}^{M}{\pi}_{m}E_{\emph m}(\tau_{c}|\Gamma)$ is the expected value of the total number of observations $\tau_c$ and $E(\tau_{s} | \Gamma )\triangleq \sum_{m=1}^{M}{\pi}_{m}E_{\emph m}(\tau_{s}|\Gamma)$ is the expected value of the total number of switchings $\tau_s$ respectively.

We adopt a Bayesian-like approach as in Chernoff's original study \cite{cd_1} by assigning a cost of $c$ for each observation, a unit loss for a wrong declaration, and $s$ be the switching cost for a single position switching.

The Bayesian risk of $\Gamma$ under hypothesis ${H}_{m}$ is given by
\begin{equation}
R_m(\Gamma) \triangleq {\alpha}_{m}(\Gamma)+cE_{m}(\tau|\Gamma)+sE_m(\tau_{s}|\Gamma).
\end{equation}

The average Bayesian risk is
\begin{equation}
\label{Bayesrisks}
{R}(\Gamma) \triangleq \sum_{m=1}^{M}{\pi}_{m}{R}_{m}(\Gamma)={P}_{e}(\Gamma)+cE(\tau|\Gamma)+sE(\tau_{s}|\Gamma).
\end{equation}

The objective is finding a policy that minimizes the Bayesian risk:

\begin{equation}
\label{objectivefunction}
\inf_\Gamma {R}(\Gamma).
\end{equation}

\subsection{Statistics}
Since only one process can be probed at a time, let $\textbf{1}_m(n)$ be the indicator function of whether process $m$ is probed at time $n$. $\textbf{1}_{m}(n)=1$ if process $m$ is probed at time $n$, and $\textbf{1}_{m}(n)=0$ otherwise. 

The log-likelihood ratio (LLR) of each process $m$ is denoted as

\begin{equation}
{l}_{m}(n) \triangleq \log\frac{g(y_m(n))}{f(y_m(n))},
\end{equation}
which reflects the goodness-of-fit of which distribution a certain region belongs to. $E_{m}(l_m(n))\triangleq D(g||f)> 0$ if the observations of region $m$ follows distribution $g(y)$, and $E_{m}(l_m(n))\triangleq -D(f||g)<0$ if the observations of process $m$ follows distribution $f(y)$, where $D(\cdot ||\cdot )$ is the KL divergence between two distributions. 

The sum log-likelihood ratio (SLLRs) of process $m$ at time $n$ is given by

\begin{equation}
{S}_{m}(n) \triangleq \sum_{t=1}^{n}{l}_{m}(t){\textbf{1}}_{m}(t),
\end{equation}
which combines the observations of the process $m$ at multiple times, and can be regarded as the score of whether the process is in an abnormal state. After a period of observations, the sum-LLRs of different processes may be different. Under $H_m$, ${S}_{m}(n)$ is a random walk with positive expected increment $E_{m}(l_m(n))=D(g||f)> 0$, while ${S}_{j}(n)$, for $j\neq m$ is a random walk with negative expected increment $E_{m}(l_j(n))=-D(f||g)< 0$.

\section{THE DBS POLICY}
\label{policy}
 
\subsection{The DBS Policy}
In this section, we propose a deterministic policy for the above active hypothesis testing problem, referred to as the DBS policy. Our goal is to detect the only one anomalous process among multiple processes. Intuitively, there are two ways to solve the detection problem. In one way, we can probe the processes that are more likely to be abnormal and confirm them one by one; In another way, we can probe the processes that are more likely to be normal and eliminate them one by one. Thus, the proposed DBS policy partitions the problem space into two cases by comparing $D(f || g)/(M-1)$ and $D(g || f)+\bigtriangleup(M)$, i.e.,
\begin{equation*}
\begin{split}
&\text{Case I: } D(g || f)+\bigtriangleup(M) \geq \frac{D(f||g)}{M-1},\\
&\text{Case II: } D( g || f)+\bigtriangleup(M) < \frac{D(f||g)}{M-1},
\end{split}
\end{equation*}
where
\begin{equation}
\label{deltaM}
\bigtriangleup(M) \triangleq \frac{s(M-2)(M+1)D(g||f)D(f||g)}{-c\log c(M-1)}
\end{equation}
is the offset value caused by the switching cost which we will explain later. It should be noted that $D(g||f)$ and $D(f||g)/(M-1)$ determine the rates at which the state of abnormal process $m$ and normal processes $j\neq m$ can be accurately inferred, where $D(g||f)$ represents the increasing slope of the test statistic $S_m(n)$ of the normal process $m$ and $D(f||g)$ represents the decreasing slope of the test statistics $S_j$ of normal processes $j\neq m$ respectively. While the observation cost is mainly determined by $D(g||f)$, $D(f||g)/(M-1)$ and $c$, the switching cost is mainly affected by $M$.  

In Case I, DBS policy probes the process which is most likely to be abnormal. The selection rule, stopping rule and decision rule are as follows:
\begin{equation}
\label{selectionrule1}
\phi (n)= m^{1}(n),
\end{equation}
\begin{equation}
\label{stopingrule1}
\tau = \min \left \{ n:S_{m^1(n)}(n) > -\log c \right \},
\end{equation}
\begin{equation}
\label{decisionrule1}
\delta = m^1(\tau),
\end{equation}
where $m^{1}(n)= \mathop{\arg\max}_{m}{S}_{m}(n)$ is the index of the process owing the highest sum-LLRs (processes with the same sum-LLRs can be ordered arbitrarily) among all the processes.

In Case II, DBS policy probes the process which is most likely to be normal and eliminates the normal processes one by one. Specifically, let $\mathcal{B}(n)$ denote the set of processes that can be declared as normal at time $n$, i.e.,
\begin{equation}
\mathcal{B}(n) = \{m: S_m(n)< \log c\}.
\end{equation}
The selection rule, stopping rule, and decision rules of DBS policy in Case II are given by
\begin{equation}
\label{selectionrule2}
\phi (n)= m^{-1}(n),
\end{equation}
\begin{equation}
\label{stopingrule2}
\tau = \min \left \{ n:\left | \mathcal{B}(n) \right | = M-1 \right \},
\end{equation}
\begin{equation}
\label{decisionrule2}
\delta = \mathcal{M} \setminus \mathcal{B}(\tau),
\end{equation}
where $m^{-1}(n)= \mathop{\arg\min}_{m\notin  \mathcal{B}(n)} {S}_{m}(n)$ is the index of the process with the lowest observation sum-LLRs among all processes that have not been declared as normal at time $n$, and $\mathcal{M}={\{ 1,2,...,M \}}$ is the set of all processes.

\subsection{Explanation}
In our policy, we define \textit{stage} as the time when a certain process is determined as abnormal or normal. There are only $1$ stage in Case I, however, the observation process will be divided into $M-1$ stages in Case II. Assume that process $m$ is in an abnormal state, i.e., $H_m$ is true. The deterministic selection rule of the DBS policy can be intuitively explained as follows. 

Since the observations $y_m(n)$ of process $m$ follow the distribution $g(y)$, the expectation of the test statistic $l_m$ is $E_m(l_m)=D(g||f)>0$. Therefore, as the number of observations increases, the sum-LLRs of process $m$ increases gradually with high probability. However, since the observations $y_j(n)$ of process $j\neq m$ follows the distribution $f(y)$, the expectation of the test statistics $l_j$ is $E_m(l_j)=-D(f||g)<0$, which means the sum-LLRs of process $j$ decreases gradually with high probability as the number of observations increases. At the beginning of the detection procedure, the sum-LLRs of all processes are zero and the decision maker randomly selects a process to probe. Subsequently, the decision maker sorts all the processes according to the sum-LLRs at each given time $n$ and selects the process $m^1(n)$ with the highest sum-LLRs. As a result, the sum-LLRs of process $m$ will gradually increases and become  $m^1(n)$ with high probability.

In Case I of DBS policy, the decision maker probes the process with the highest sum-LLRs at each given time. The test is finalized once sufficient information is gathered from the process, when the highest sum-LLRs exceeds the threshold of $-\log c$. Therefore, the asymptotic detection time approaches $- \log c/D(g||f)$ since the abnormal process can be probed with higher probability than other processes at each time. In this case, the number of switching is bounded because the abnormal process will quickly become the $m^{1}(n)$ process with high  probability. In addition, the probability of switching decreases as the time $n$ increases, since the gap between $m^1(n)$ and $m^2(n)$ increases as $n$ increases.

In Case II, the decision maker eliminates normal processes one by one and identifies the abnormal process finally. As a result, the detection procedure will be divided into $M-1$ stages. At each stage, the process $m^{-1}(n)$ which is most likely to be normal is probed at each time and the asymptotic observation time of each stage approaches $\frac{-\log c}{D(f||g)}$. The test is finalized once $\left | \mathcal{B}(n) \right | = M-1$. Thus the asymptotic observation time of entire detection procedure approaches  $\frac{-(M-1)\log c}{D(f||g)}$.

With the goal of minimizing the objective function (\ref{objectivefunction}), the selection rules of Case I and Case II are established as (\ref{deltaM}). Considering the impact of switching on the strategy, we introduced an offset $\bigtriangleup(M)$ since there will be more switchings in Case II which has $M-2$ more stages than Case I. It is worthy noting that the value of $\bigtriangleup(M)$ changes as the sample cost $c$, the single-switching cost $s$ or the number of processes  $M$ changes.  On the one hand, $\bigtriangleup(M)$ increases as $\frac{s}{-c\log c}$ increases when $M$ is fixed, which means that as $\frac{s}{-c\log c}$ increases, the decision maker is more inclined to choose Case I. The reason is that the effect of switching cost on the objective function increases as $\frac{s}{-c\log c}$ increases, which leads the decision maker to choose Case I with fewer switching times. On the other hand, $\bigtriangleup(M)$ increases as $M$ increases  when $\frac{s}{-c\log c}$ is fixed, which means that it's more inclined to choose Case I as $M$ increases, since there will be more stages in Case II as $M$ increases, and switching will occur at each stage which leads to an increase in switching cost.


\begin{figure}
	\centering
	\includegraphics[width=0.8\linewidth]{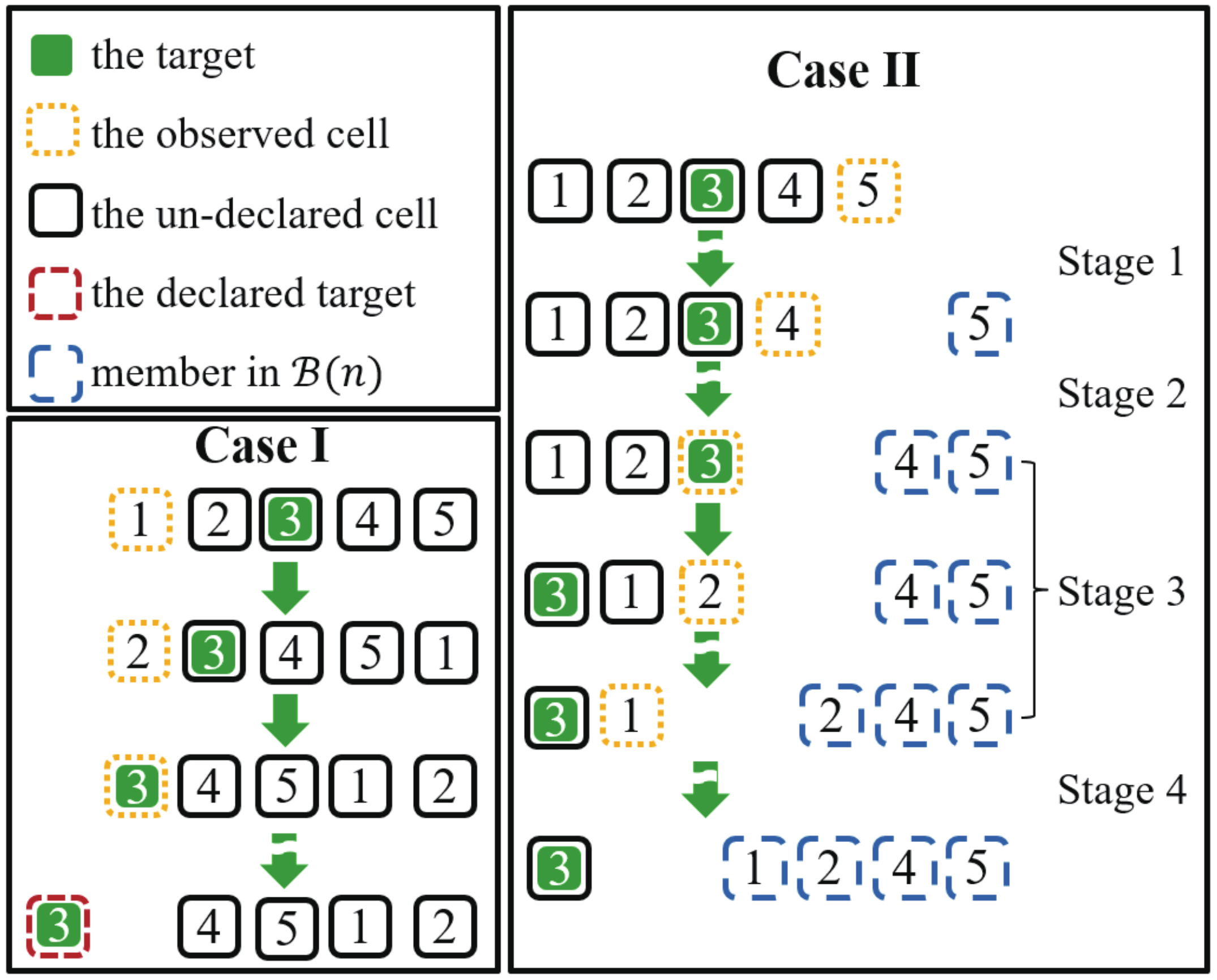}
	\caption{Illustration the two cases of DBS Policy ($M=5$).}
\label{lb1Numberofobservations}
\end{figure}

\subsection{Example}
We illustrate the two cases of DBS policy in Fig.~\ref{lb1Numberofobservations} with an example that find the only abnormal process among $M=5$ processes to clarify our idea.

Fig.~\ref{lb1Numberofobservations} simulates the specifics of the DBS policy for Case I and Case II, respectively, when process 3 is an anomalous process. For Case I, the decision maker probes the process $m^1(n)$ with the highest sum-LLR. At the beginning of the detection procedure, since sum-LLRs of all processes are zero, a process is randomly selected to probe at $t=1$. Suppose the decision maker selects process 1. Since process 1 is in a normal state, the sum-LLR of process 1 will decreases, so at $t=2$, the order of sum-LLRs of all processes becomes $\{2,3,4,5,1\}$. Select process 2 to probe. In the same way, the sum-LLR of process 2 will also decrease, and the order of sum-LLRs of all processes may become $\{3,4,5,1,2\}$. Assume that the decision maker selects process 3 to probe at $t=3$. Since process 3 is in an abnormal state, the sum-LLR of process 3 increases and ranks first, i.e., $m^1(n)=3$ and $S_{m^1(n)}(n)>0$. From then on, the decision maker will continue to probe process 3 with high  probability until $S_{m^1(n)}(\tau)\geq -\log c$. Then the decision maker will stop the detection procedure and declares process 3 to be an anomaly.

For Case II, the decision maker probes process $m^{-1}(n)$ at each given time. Suppose the process 5 is randomly selected to probe at the beginning of the detection procedure. Since process 5 is in a normal state, the sum-LLR of process 5 will decreases and be ranked last. Thereafter, the decision maker will continue to probe process 5 until $S_{m^{-1}(n)}(n)<\log c$, and then stop taking observations from process 5 and declare it as normal. Repeat this operation until all processes in normal state are declared as normal.

\section{Performance Analysis and Comparison}
\label{analyze}
In this section, we will analyze the performance of the DBS policy and compare DBS policy with DBS policy and R-SPRT policy.

\subsection{Performance Analysis}
\label{DBS-optimal}
In this section, we analyze the performance of the DBS policy with different $s$ and $c$. Theorem 1 and Theorem 2 are presented to prove that the DBS policy is optimal in terms of minimizing the objective function as the observation cost $c$ approaches zero.

It is obvious that the order difference between $s$ and $c$ will affect the choice of strategy and the performance of the corresponding policy. Considering the order relationship between the observation cost and switching cost, the problem is divided into the following two scenarios:

\textbf{Scenario 1}: the single-switching cost $s$ becomes insignificant relative to $-c\log c$ in the asymptotic regime of $c\rightarrow 0$, i.e., $\lim_{c\rightarrow 0} \frac{s}{-c\log c}=0$. 

\textbf{Scenario 2}: the single-switching cost $s$ is bounded below by $-c\log c$ in the asymptotic regime of $c\rightarrow 0$, i.e., $\lim_{c\rightarrow 0} \frac{s}{-c\log c} >0$.

According to the subsequent analysis, it is clear that the observation time $E(\tau_c)$ of DBS policy satisfies $E(\tau_c)=\frac{-\log c}{I^*(M)}$ as $c\rightarrow 0$. Thus, the observation cost can be denoted as $cE(\tau_c)=\frac{-c\log c}{I^*(M)}$, which is a function related to $-c\log c$. Meanwhile, the number of switchings is limited and the total switching cost can be denoted as a function related to $s$. Therefore, in order to better analyze the performance of the DBS policy under the objective function in the form of Bayesian risk, the discussion is based on the order of $s$ and $-c\log c$. Scenario 1 means that the total switching cost is much smaller than the total observation cost and is negligible compared with the total observation cost. In Scenario 2 the total switching cost and total observation cost are comparable.

We will analyze and prove the optimality of the DBS policy in two scenarios as $c\rightarrow 0$, respectively. The following theorems show that DBS policy is asymptotically optimal or order optimal regarding minimizing the Bayesian risk as $c$ approaches zero. Theorem 1 will show that the Bayesian risk of DBS policy will be infinitely close to the lower bound of Bayesian risk if $s=o(-c\log c)$ as $c\rightarrow 0$. Theorem 2 will show that the Bayesian risk of the DBS policy is higher than the lower bound of the Bayesian risk, but the ratio of the two is bounded.

\begin{theorem}[Asymptotic Optimality of DBS Policy]
\label{theorem1}
Let $R(\text{DBS})$ and $R(\Gamma)$ be the Bayesian risks under the DBS policy and any other policy $\Gamma$, respectively. If $s=o(-c\log c)$, then, 
\begin{equation}
R(\text{DBS}) \sim \frac{-c\log c}{I^*(M)} \sim \inf_{\Gamma} R(\Gamma) \text{ as } c\rightarrow 0,
\end{equation}
where,
\begin{equation*}
\label{imk}
I^*(M) = \begin{cases}
D(g||f), & \text{ if Case I},\\ 
D(f||g)/(M-1), & \text{ if Case II}.        
\end{cases}
\end{equation*}
\end{theorem}

\begin{IEEEproof}
	For a detailed proof see Appendix A. We provide here a sketch. In the first step, we prove that the objective function based on Bayesian risk has an asymptotic lower bound $\inf_{\Gamma} R(\Gamma)$ and give the form of the asymptotic lower bound. Then, we show that the Bayesian risk $R$ under DBS policy approaches the asymptotic lower bound as $c\rightarrow 0$. 
	
	Based on the three perspectives of error probability, sampling cost and switching cost, the asymptotic behavior of Bayesian risk is established. In Lemma 1, we show that the asymptotic lower bound on the total switching cost is $\sum_{k}(k-1)\pi_k$, the asymptotic lower bound on the sampling cost is $\frac{-c\log c}{I^*(M)}$, and the error probability is $O(c)$ following the same proof idea as in \cite{cohen_bayes}.
	
	The basic idea of proving that the DBS policy is asymptotic optimal is to prove that the Bayesian risk of the DBS policy approaches the asymptotic lower bound as $c$ approaches 0. In App.1, we prove that the asymptotic expected detection time approaches $\frac{-\log c}{I^*(M)}$ while the error probability is $O(c)$. In addition, we prove that the number of switching has an upper bound which is independent of $c$. 
	
	Then, we analyze the expected number of observations and expected switching times of the DBS policy. Take the proof of Case II as an example, we need split the testing into $M-1$ stage. Each stage is defined according to the time when a certain process is placed into set $\mathcal{B}$. Then we analysis the three last passage times at every stage $k$, denote as $\tau_1^k$, $\tau_2^k$, $\tau_3^k$, where $\tau_1^k$ is the time when region $m^{-1}(n)$ is always normal for all $n\geq \tau_{1}^k$, $\tau_2^k$ is the time when region $m^{-1}(n)$ no longer changes for all $n\geq \tau_{2}^k$, and $\tau_3^k$ is the time when the region $m^{-1}(n)$ is placed into $\mathcal{B}$. The switching only occurs during $\tau_1^k$ and $\tau_2^k$, so the key point in our policy is to prove that there exist constants $C>0$ and $\gamma>0$ such that $\textbf{P}_m(\tau_{1}^{k}>n)\leq Ce^{-\gamma n}$ and $\textbf{P}_m(\tau_{2}^{k}>n)\leq Ce^{-\gamma n}$. Therefore, we can get the upper bound of the expected switching times as $c\rightarrow 0$, and denote $d$ as the upper bound. In addition, combining the above conclusion and the property of SPRT, we can draw the conclusion that $E(\tau_3^k)\sim -\log c/D(f||g)$ in Case II, so the asymptotic expected detection time approaches $-\log c/I^*(M)$. 
	
	Last but not Least, we will analyze the optimality of the DBS policy as $c$ approaches zero based on the relationship between $s$ and $c$. Since $\lim_{c\rightarrow 0} \frac{s}{-c\log c}=0$, which means $s$ is much smaller than $-c\log c$, the ratio of expected number of switching to the number of observations $\frac{s\cdot D\cdot I^*(M)}{-c\log c}$ approaches zero as $c\rightarrow 0$. Thus, the ratio of asymptotic lower bound of Bayesian risk to the Bayesian risk of the DBS policy $\frac{R(\text{DBS})}{\inf_{\Gamma} R(\Gamma)}$ is equal to 1 as $c\rightarrow 0$, which means the DBS policy is asymptotic optimal as $c\rightarrow 0$.	
\end{IEEEproof}

\begin{theorem}[Order Optimality of DBS Policy]
	\label{theorem2}
	Let $R(\text{DBS})$ and $R(\Gamma)$ be the Bayesian risks under the DBS policy and any other policy $\Gamma$, respectively. If $s=\Omega (-c\log c)$, then, \\
	\begin{equation}
	R(\text{DBS})=O(\inf_{\Gamma} R(\Gamma))\text{ as } c\rightarrow 0 
	\end{equation}
\end{theorem}

\begin{IEEEproof}
	For a detailed proof see Appendix B. Theorem 2 can be proved in a similar way to theorem 1. Theorem 2 has the same expected switching time and expected observation time as theorem 1. The difference between Theorem 2 and Theorem 1 is that the relationship between s and c is different. For Theorem 2, since $\lim_{c\rightarrow 0} \frac{s}{-c\log c}>0$, the ratio of asymptotic lower bound of Bayesian risk to the Bayesian risk of the DBS policy $\frac{R(\text{DBS})}{\inf_{\Gamma} R(\Gamma)}$ is larger than 1 but bounded by positive infinity as $c\rightarrow 0$ , which means the DBS policy is order optimal as $c \rightarrow 0$.
\end{IEEEproof}

In the following, we will analyze the DGF policy and the R-SPRT policy when they are applied to the anomaly detection problem considering switching cost. The DGF policy has been shown to be an asymptotically optimal policy that minimizes the detection time subject to an error probability constraint. And the R-SPRT policy is an optimal policy that minimizes the number of switchings.

\subsection{Comparison with the DGF Policy}
\label{DGF}
As DBS policy is inspired by DGF policy in \cite{cohen_bayes}, we would like to highlight the difference between DBS policy and DGF policy to explain how the Bayesian risk considering switching cost is reduced in DBS policy.

\subsubsection{The DGF Policy}

The DGF policy is a  deterministic policy. At each time, the selection rule $\phi(n)$ of the DGF policy chooses cells according to the order of their sum-LLRs which is similar to the DBS policy. Specifically, based on the relative order of $D(g||f)$ and $D(f||g)/(M-1)$, either the process with the highest sum-LLRs or the process with the second highest sum-LLRs are chosen, i.e., 

\begin{equation*}
\phi(n) = \begin{cases}
m^1(n), & \text{ if } D(g||f)\geq \frac{D(f||g)}{M-1},\\ 
m^2(n), & \text{ if } D(g||f)< \frac{D(f||g)}{M-1}. 
\end{cases} 
\end{equation*}

The stopping rule and decision rule under the DGF policy are given by:
\begin{equation*}
\tau = \inf \{n:m^1(n)-m^2(n) \geq -\log c\},
\end{equation*}
and,
\begin{equation*}
\delta=m^1(n).
\end{equation*}

\subsubsection{Comparison}
In Case I, both the DBS policy and the DGF policy are to probe the process which is most likely to be abnormal until the decision makers declare one process as abnormal. The selection rule of DGF policy is to probe the $m^1(n)$ process at each time, which is the same as DBS policy. The stopping rule under DGF policy is associate with the difference between the highest observed sum-LLRs $S_{m^1(n)}(n)$ and the second highest observed sum-LLRs $S_{m^2(n)}(n)$. Once $S_{m^1(n)}(n)-S_{m^2(n)}(n)\geq -\log c $, the decision maker of DGF policy finalizes the detection procedure. Meanwhile, the decision maker of DBS policy stops the detection procedure once the highest observed sum-LLRs $S_{m^1(n)}(n)$ is greater than $-\log c$. Since there is only one anomalous process, the second highest observed sum-LLRs is less than 0 with high  probability. Because of the setting of the selection rule and stopping rule, DGF policy has fewer excepted observation times than DGF policy while it has the same expected switching times as DBS policy, which makes the Bayesian risk of DGF policy slightly lower than DBS policy. However, the Bayesian risk gap between DGF and DBS policy caused by the number of observations is small enough to not affect the asymptotic optimality of the DGF policy. 

However, in Case II, DGF policy probes the process with the second sum-LLRs at each time, and the testing is terminated once the difference between the highest and the second highest observed sum-LLRs is greater than $- \log c$. Since the sum-LLRs of probed process $m^{2}(n)$ at time $n$ will decrease with high probability after current observation, the probed process will always changes, which cause a large switching cost. In contrast, DBS policy always probe the process $m^{-1}(n)$ in Case II, since the observed sum-LLRs of process $m^{-1}(n)$ at time $n$ will decrease with high probability after current probing, DBS policy will probe the same process with high probability until it has been declared. Thus, the number of switching times is reduced.

\subsection{Comparison with the R-SPRT Policy}
\label{SPRT}
In this part, we will compare the difference between the DBS policy and the R-SPRT policy, which has the smallest expected switching time.

\subsubsection{The R-SPRT Policy}
The R-SPRT policy has a random selection rule, where a series of SPRTs are performed until all the abnormal process is tested in a random order. In other words, the decision maker selects one process to probe randomly and performs SPRT test for the currently probed process until its state is declared. Once one process is declared as abnormal, the detection procedure will be finished.

In R-SPRT, the stopping and decision rules are given by comparing the sum-LLRs $S_m(n)$ with boundary values at each time $n$. Thus, the R-SPRT test is carried out as follows:
\begin{itemize}
	\item If $S_m(n)\in (\log c,-\log c)$, continue to take observations from process $m$.
	\item If $S_m(n)\leq -\log c$, stop taking observations from process $m$ and declare it as normal. In addition, select another process whose state hasn't been declared randomly to probe at the next time.	
	\item If $S_m(n)\geq -\log c$, stop taking observations from process $m$ and declare it as abnormal. Then, finish the whole detection procedure and $\tau=n$.
\end{itemize}

Since the R-SPRT policy selects processes randomly, each process has same probability to be selected at each stage. The expected switching time of the R-SPRT policy can be given by
\begin{equation*}
E(\tau_s|\text{R-SPRT})=\frac{M-1}{2},
\end{equation*}
while the expected observation time of the R-SPRT is
\begin{equation*}
E(\tau_c|\text{R-SPRT})=-\log c\left(\frac{1}{ D(g||f) } +\frac{M-1}{2D(f||g)} \right).
\end{equation*}

\subsubsection{Comparison}
In this part, we will compare the performance of the R-SPRT policy and the DBS policy on the objective function (\ref{objectivefunction}). The Bayesian risks of the R-SPRT policy and the DBS policy can be given by:

\begin{equation*}
\begin{aligned}
R(\text{\text{R-SPRT}}) & =c E(\tau_c|\text{R-SPRT}) + sE(\tau_s|\text{R-SPRT}) \\
&= -c \log c \left(\frac{1}{ D(g||f) } +\frac{M-1}{2D(f||g)} \right) + s\frac{M-1}{2}, \\
R(\text{DBS}) & =c\cdot E(\tau_c|\text{DBS}) + s\cdot E(\tau_s|\text{DBS}) \\
&= -c \log c \frac{1}{I^*(M)}  + sE(\tau_s|\text{DBS}),
\end{aligned}
\end{equation*}
where there exist $\frac{M-1}{2}<D<\infty$ such that $\frac{M-1}{2} < E(\tau_s|\text{DBS}) < D$.

Assume that $R(\text{DBS})$ is lower than $R(R-SRPT)$, i.e., 
\begin{equation*}
\begin{split}
R(\text{DBS}) & \leq -c \log c \frac{1}{I^*(M)} + sD \\
& < -c \log c \left(\frac{1}{ D(g||f) } +\frac{M-1}{2D(f||g)} \right) + s\frac{M-1}{2} \\
& = R(\text{R-SPRT}),
\end{split}
\end{equation*}

we can draw the conclusion as follow:
\begin{equation*}
\frac{s}{-c\log c} <\frac{ \frac{1}{D(g||f)} + \frac{M-1}{2D(f||g)} - \frac{1}{I^*(M)} } { D-\frac{M-1}{2} }.
\end{equation*}

In another words, when the order relationship between $s$ and $-c\log c$ meets the above conditions, the DBS performs better than the R-SPRT policy.

\section{NUMERICAL RESULTS}
\label{result}
\label{simulation}
In this section, we present numerical examples to illustrate the performance of DBS policy. We consider the case where there is only one anomalous process among $M$ processes. The DBS policy will be compared with Chernoff test\cite{cd_1}, DGF policy\cite{cohen_bayes}, Sluggish policy\cite{cohen_bayes} and R-SPRT policy. 

Chernoff test is a randomized policy, which works as below for anomaly detection problem in this case. When $D(g||f)\geq D(f||g)/(M-1)$, Chernoff policy selects the $m^1(n)$ process at each given time since there is only one abnormal process and the $m^1(n)$ is most likely to be abnormal. When $D(g||f)< D(f||g)/(M-1)$, the decision maker will select one process randomly with equal probability from processes $m\neq m^1(n)$. As a result, the Chernoff test has a different selection rule with the DBS policy.

Sluggish policy is an extended algorithm based on Chernoff test, which introduce costs for switching of actions. The Sluggish policy merely slows down the switching of actions via an independent and identically distributed Bernoulli modulation process, which performs well in the active hypothesis testing problem considering switching cost.

We set the simulations with the following parameters. When process $m$ is probed at time $n$, an observation $y_m(n)$ is independently drawn from a Poisson distribution $f\sim \text{Pois}(\lambda_f)$ or $g\sim \text{Pois}(\lambda_g)$, depending on whether the process is the target or normal. It is easy to show that

\begin{equation}
\begin{split}
D(g||f)={{\lambda }_{f}}-{{\lambda }_{g}}+{{\lambda }_{g}}\log (\frac{{{\lambda }_{g}}}{{{\lambda }_{f}}}),\\
D(f||g)={{\lambda }_{g}}-{{\lambda }_{f}}+{{\lambda }_{f}}\log (\frac{{{\lambda }_{f}}}{{{\lambda }_{g}}}).
\end{split}
\end{equation}

Let $T_c(\Gamma)$ be the number of observations under the policy $\Gamma$, and $T_c^{LB}=\frac{-\log c}{I^*(M)}$ be the asymptotic lower bound on the observation times as $c \rightarrow 0$. Define 
\begin{equation*}
T_c'(\Gamma)\triangleq (T_c(\Gamma)-T_c^{LB})/T_c^{LB}
\end{equation*}   
as the relative ratio of observations under policy $\Gamma$ compared to the asymptotic lower bound of observation.

Let $R(\Gamma)$ be the Bayesian risks under the policy $\Gamma$, and $R_{LB}=\frac{-c \log c}{I^*(M)} + s\cdot \sum_{k}(k-1)\pi_k$ be the asymptotic lower bound on the Bayesian risk as $c\rightarrow 0$. Define
\begin{equation}
L(\Gamma)\triangleq (R(\Gamma)-R_{LB})/R_{LB}
\end{equation}
as the Bayesian relative risk under policy $\Gamma$ compared to the asymptotic lower bound, which serve as performance measures of the tests in the finite regime. Following Theorems 1,2, we expect $L(\text{DBS})$ to approach 0 when $s=o(-c\log c)$as $c\rightarrow 0$ and is bounded by positive infinity when $s=\Omega(-c \log c)$.

\subsection{$s/c$ is fixed}

\begin{figure}
	\centering
	\subfloat[The relative ratio of observations]{ 
		\includegraphics[width=0.5\linewidth]{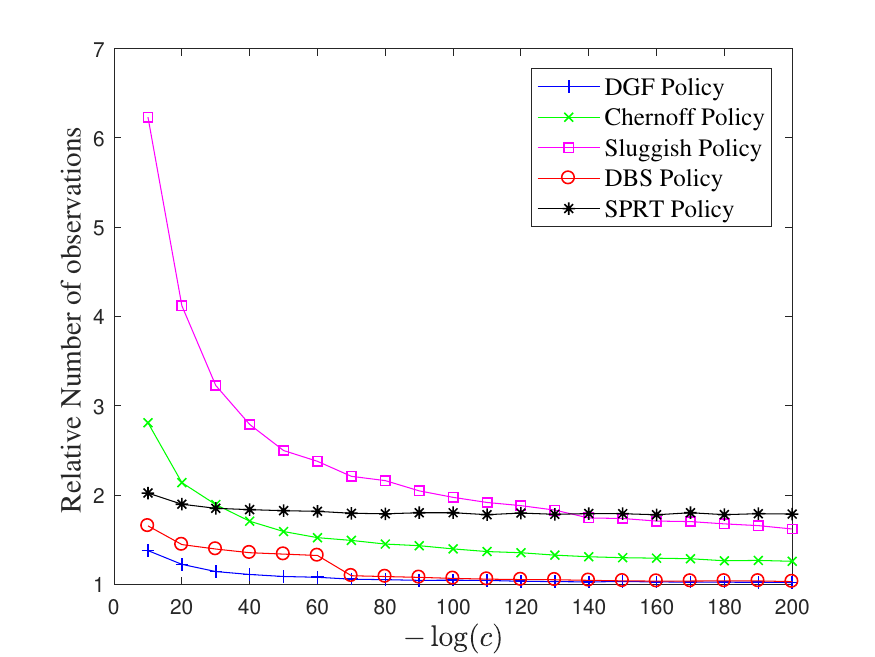} } \\
	\subfloat[Switching number]{ 
		\includegraphics[width=0.5\linewidth]{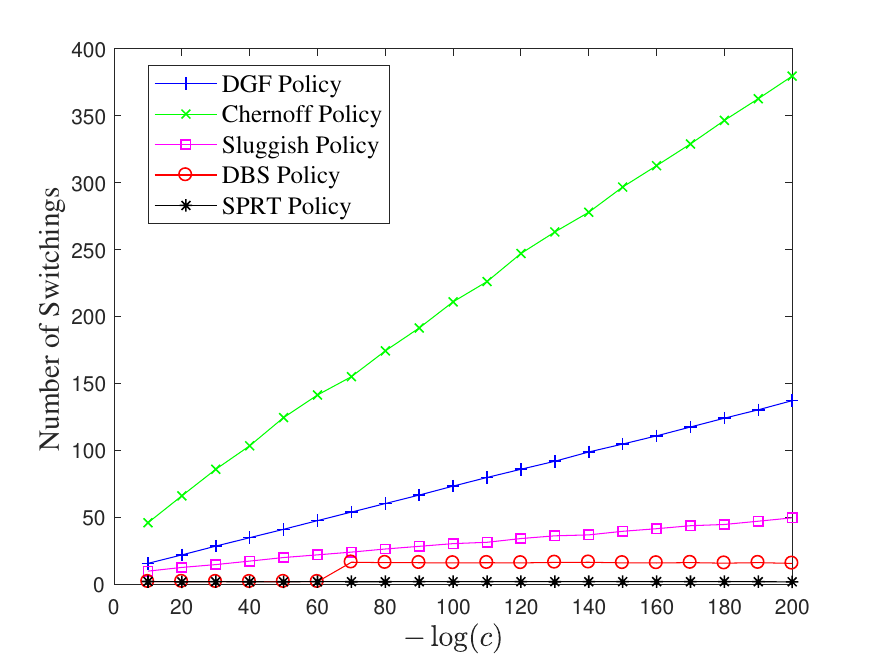} } \\
	\subfloat[Bayes relative risk]{ 
		\includegraphics[width=0.5\linewidth]{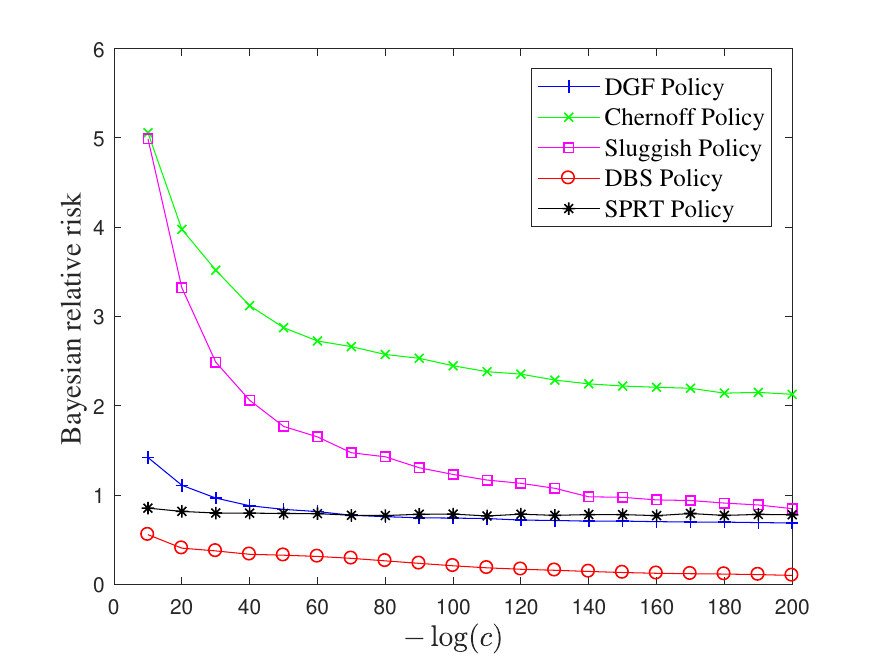} }
	\centering
	\caption{DBS Policy in $M=5$, $s=2c$, $\lambda_f = 0.4$, $\lambda_g = 0.001$}
\label{s/cisfixed}
\end{figure}


First we consider the case where $M=5$ and $s=2c$, which means $\frac{s}{-c\log c} = 0$ as $c \rightarrow 0$. In this case, the loss of a single switch and a single observation is comparable, while the total switching cost is much smaller than the total observation cost. For Sluggish policy\cite{cohen_bayes}, the switching probability is set to $p=0.1$, which means switching occurs approximately every ten observations. We set $\lambda_f=0.4$, $\lambda_g=0.001$ and obtained $D(g||f)\approx 0.39$, $D(f||g)/(M-1) \approx 0.50$. The performance of all algorithms is presented in Fig. \ref{s/cisfixed}, which are the average of 1000 trials. 

It is shown that Sluggish policy has the largest number of observations and DGF policy is the smallest in  Fig. \ref{s/cisfixed}(a). In Fig. \ref{s/cisfixed}(b), we can find that the switching number of R-SPRT policy is less than other policies. Regardless of the relative ratio of observation as Fig. \ref{s/cisfixed}(a) or the number of switching as Fig. \ref{s/cisfixed}(b), the performance of the DBS strategy is both  sub-optimal. The dashed rectangular area of Fig. \ref{s/cisfixed}(a) and Fig. \ref{s/cisfixed}(b) shows that the observation delay and switching number of DBS policy change suddenly around $-\log c=70$. 
The reason is that DBS policy chooses to probe the process with the largest sum-LLRs since  it is in Case I when $-\log c<70$. However, because of the change of $c$, DBS policy changes to Case II and chooses to observe the $M-1$ normal processes when $-\log c > 70$. The transition between cases increases the number of stages, leading to an increase in the number of switchings and a decrease in the number of observations.As a result, the Bayes relative risks  will continuously decrease and approach to $0$ as $c\rightarrow 0$.

\subsection{$s=o(-c\log c)$}

In this section, we consider the scenario where $\frac{s}{-c\log c}=5c$, which means $\frac{s}{-c \log c}=0$ as $c \rightarrow 0$. In this case, total switching cost is sufficiently small in comparison with the total observation cost. In addition, $\triangle(M)$ approaches 0 as $c \rightarrow 0$, so Case I and Case II of DBS policy can be simplified to the following form,

\begin{equation*}
\begin{split}
& \text{Case I}: D(g||f) \geq \frac{D(f||g)}{M-1}, \\
& \text{Case II}: D(g||f) < \frac{D(f||g)}{M-1}. 
\end{split}
\end{equation*}

\begin{figure}[htbp]
	\centering
	\subfloat[The relative ratio of observations]{ 
		\includegraphics[width=0.5\linewidth]{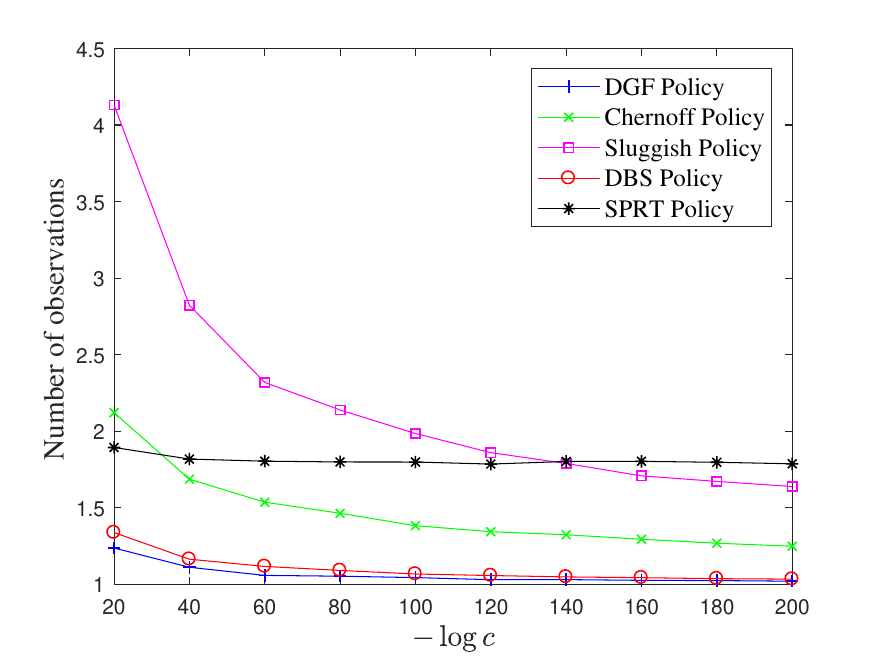} } \\
	\subfloat[Switching number]{ 
		\includegraphics[width=0.5\linewidth]{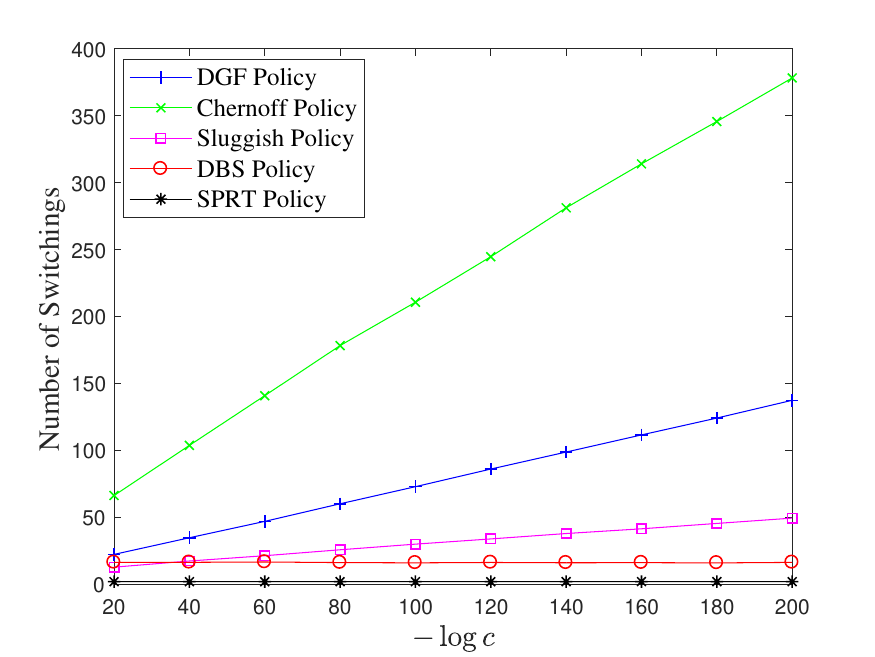} } \\
	\subfloat[Bayes relative risk]{ 
		\includegraphics[width=0.5\linewidth]{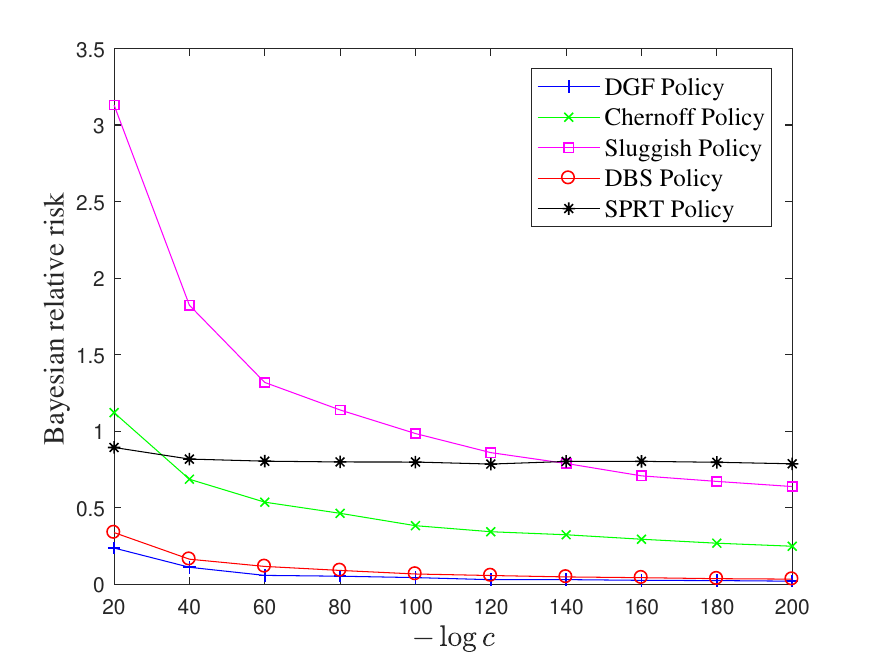} }
	\centering
	\caption{DBS Policy in $M=5$, $\frac{s}{-c\log c}=5c$, $\lambda_f = 0.001$, $\lambda_g = 0.4$}
\label{oc1}
\end{figure}


We set $\lambda_f=0.001, \lambda_g=0.4$ and obtained $D(g||f)\approx 0.39, D(f||g)/(M-1)\approx 0.50$, when DBS policy, DGF policy and Chernoff policy are all in Case II and inclined to probe the normal processes. In this case, DBS policy selects process $m^{-1}(n)$ at each given time $n$, while the decision maker of DGF policy selects process $m^2(n)$ at each given time. And the Chernoff test draws one process randomly with equal probability from processes $\{ m^2(n), m^3(n),...,m^M(n) \}$ at each given time $n$.  Since the test statistics from normal processes are less than zero with high probability, the LLR of normal process will be lower and lower. Thus, the decision maker of DBS policy will probe the same process at each given time with high probability, while the decision maker of DGF policy switches the observed process constantly. The performance of all algorithms is presented in Fig. \ref{oc1}.

\subsection{$s=\Omega(-c\log c)$}

\begin{figure}[htbp]
	\centering
	\subfloat[The relative ratio of observations]{ 
		\includegraphics[width=0.5\linewidth]{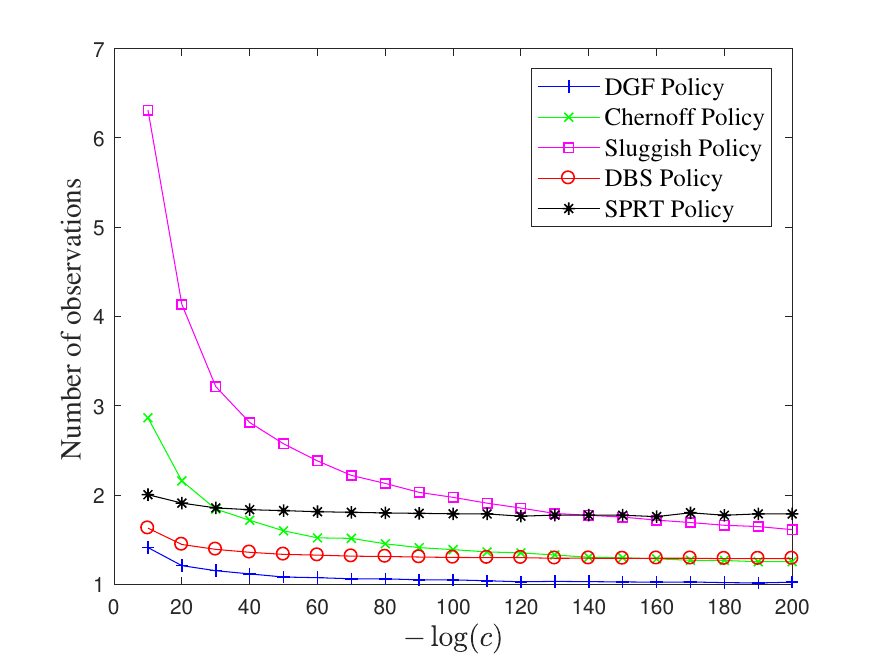} } \\
	\subfloat[Switching number]{ 
		\includegraphics[width=0.5\linewidth]{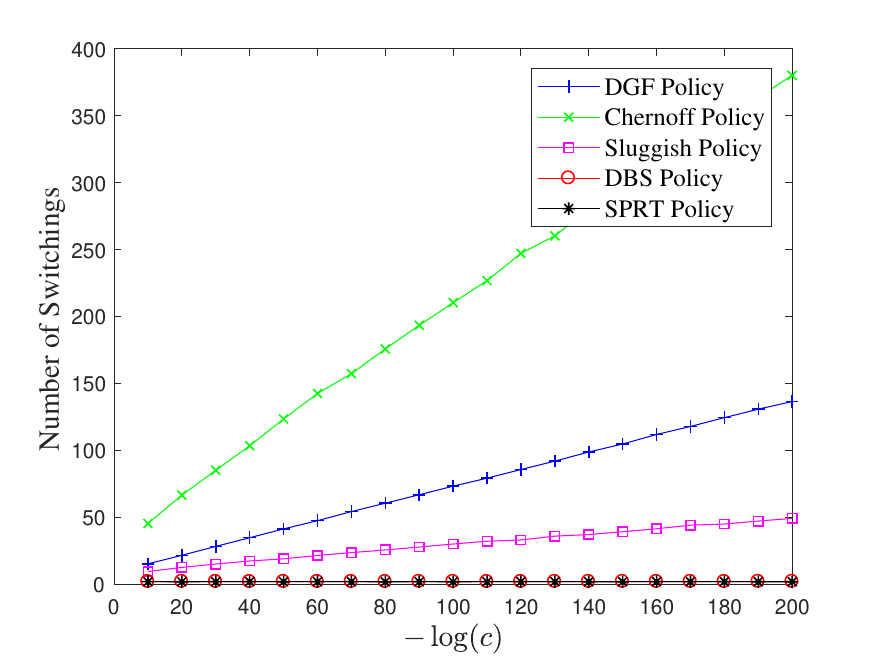} } \\
	\subfloat[Bayes relative risk]{ 
		\includegraphics[width=0.5\linewidth]{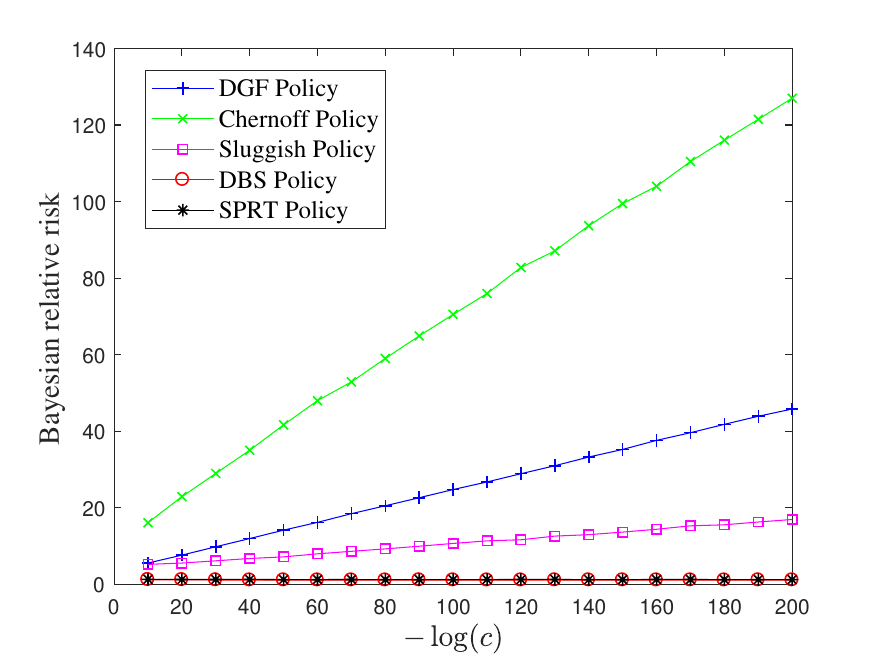} }
	\centering
	\caption{DBS Policy in $M=5$, $\frac{s}{-c\log c}=5c$, $\lambda_f = 0.001$, $\lambda_g = 0.4$}
	\label{omegac}
\end{figure}


Then we consider the case where $\frac{s}{-c\log c}=2$, which means the single-switching cost $s$ is large enough that the cost caused by switching cannot be ignored. In addition, $\triangle(M)$ does not change with the change of $c$. We set $\lambda_f=0.4$, $\lambda_g=0.001$ and obtained $D(g||f)\approx 0.39$, $D(f||g)/(M-1) \approx 0.50$, $D(g||f)+\triangle(M) \geq 7$, DBS policy is in Case I for all values of $c$ while DGF policy is in Case II. The performance of all algorithms is presented in Fig. \ref{omegac}. It is shown that DGF policy has less observations, but DBS policy has less switching. Fig. \ref{omegac}(c) shows that DBS policy is optimal among all algorithms.

\section{Conclusion}
\label{conclusion}

The problem of detecting an anomalous process among multiple processes considering switching cost is studied. Due to the resource constraints, only one process can be probed at each time, and the observations from each process follow two different distributions, depending on whether the process is normal or abnormal. Each swithcing across processes incurs an additional switching cost $s$ and each observation incurs detection delay $c$, while a wrong declaration incurs a loss of $1$. The objective is to find a search strategy that minimizes the expected detection time and the expected switching cost subject to error probability constraint. Considering the order relationship between the detection delay and switching cost, we divided the problem into two scenarios: the total switching cost is much smaller than the detection delay(i.e., $s=o(-c\log c)$ as $c\rightarrow 0$), or the total switching cost is comparable with the detection delay(i.e., $s=\Omega(-c\log c)$ as $c\rightarrow 0$). We propose a deterministic policy and proved that our policy is asymptotically optimal when the total switching cost is much smaller than the total observation cost and is order optimal when the total switching cost is comparable with the observation cost as $c$ approaches zero. We also offers better performance in the finite regime for different scenarios respectively. Our policy can also be applied to target detection, fraud detection and other read-world scenarios.

Deriving optimal policies for the anomaly detection problem with switching cost considered in this paper requires the assumption that only one process is in an abnormal state and the decision maker can probe only one process at each time. In the future, we will further extend the problem to handle the case where multiple anomalous processes are present and multiple processes can be probed simultaneously. In this paper, we also assume that the observation distributions under both hypothesis are completely known. In further research, we expect that the policy can be extended to the situation where the distribution of each process is heterogeneous or the number of processes is unknown.

\appendix

\subsection{The Proof of Theorem 1}

In this section, we will prove the asymptotic optimality of DBS policy in the situation where $s=o(-c\log c)$ as $c$ approaches $0$. In App A.1, we show that $\frac{-c \log c}{I^*(M)}+s\cdot \sum_{k=1}^{M}(k-1)\pi_k$ is an asymptotic lower bound on the Bayesian risk. Then, we show in App A.2 that if $s=o(-c\log c)$, the Bayesian risk under DBS policy $R(\text{DBS})$ approaches the asymptotic lower bound as $c\rightarrow 0$.

\textit{1) The Asymptotic Lower Bound of the Bayesian risk}

\begin{proposition}
	\label{proposition2-1}
	Let $R(\Gamma)$ be the Bayesian risks under any policy $\Gamma $. Then,
	\begin{equation}
	R(\Gamma) \geq  c\cdot \frac{-\log c}{I^*(M)} + s\cdot \sum_{k=1}^{M} (k-1)\pi_k^\prime \text{ as } c \rightarrow 0,
	\end{equation}
	where $\pi_k^\prime$ denote the $k^{th}$ highest probability of the priori probability set $\{\pi_m \}_{m=1}^{M}$ and
	\begin{equation}
	\label{imk}
	I^*(M) = \begin{cases}
	D(g||f), & \text{ if Case I},\\ 
	D(f||g)/(M-1), & \text{ if Case II}.  \\       \end{cases}
	\end{equation}
	
	\begin{IEEEproof}
	We divide the discussion about the asymptotic lower bound of Bayesian risk into the following three part: the probability of error, the expected detection cost and the expected switching cost.
	
	In App.A.1 of DGF Policy\cite{cohen_bayes}, it has been proved that $ \frac{-c \log c}{I^*(M)}$ is an asymptotic lower bound on the Bayesian risk without switching cost, which can be regarded as the asymptotic lower bound on the summation of the probability of error and the expected detection cost, i.e.,
	\begin{equation}
	P_e(\Gamma)+cE(\tau_c|\Gamma) \geq  c\cdot \frac{-\log c}{I^*(M)} \text{ as } c \rightarrow 0.
	\end{equation}
		
	Then, we will analyze the lower bound of the expected switching time. Assume that each process must be probed continuously until its state is declared. In another words, switching across processes is allowed only when the state of the currently probed process is declared, which is the minimal number of switchings during the detection procedure. Thus at most $M-1$ switchings will occur and the expected switching time can be denoted as
	\begin{equation*}
	E(\tau_s)=\sum_{t=0}^{M-1} tP(\tau_s=t),
	\end{equation*}
	where $\sum_{t} P(\tau_s=t) =1$.
	
	Sorting $\{\pi_m \}_{m=1}^{M}$ in descending order, we get the set $\{\pi_k^\prime \}_{k=1}^{M}$, where $\pi_1^\prime \geq \pi_2^\prime \geq \cdots \geq \pi_k^\prime \geq \cdots \geq \pi_M^\prime$ and  $\pi_k^\prime$ denote the $k$-th highest probability of the priori probability set $\{\pi_m \}_{m=1}^{M}$. Let $\pi_{ik}^\prime$ be the corresponding posterior probability at the $i$-th choice. Let $p_k^i$ be the probability that process $k$ is selected at the $i$-th choice. Then the probability of selecting the abnormal process at the $t$-th choice can be expressed as
	\begin{equation*}
	P(\tau_s=t)=(1-P(\tau_s< t))\sum_{k=1}^M \pi_{tk}^\prime \cdot p_k^t. 
	\end{equation*}
	It is known that $P(\tau_s=t)$ can be expressed as an expression independent of $t$. Thus, the problem of minimizing $E(\tau_s)$ can be transformed into maximizing $P(\tau_s=t)$ for each $t$.
	
	Then the probability of selecting the abnormal process at the first choice, which means the number of switching is equal to zero, is 
	\begin{equation*}
	P(\tau_s=0)=\sum_{k=1}^M \pi_{0k}^\prime \cdot p_k^1, 
	\end{equation*}  
	where $\pi_{0k}^\prime=\pi_{k}^\prime$. The solution for maximizing $P(\tau_s=0)$ is to make $p_1^1$ be $1$ and the rest to be $0$. Thus, $ P(\tau_s=0) \leq \pi_1^\prime$. 
	
	Then the probablity of selecting the abnormal process at the second choice is 
	\begin{equation*}
	\begin{split}
		P(\tau_s=1) & = (1-P(\tau_s<1))\sum_{k=1}^M \pi_{tk}^\prime \cdot p_k^t \\
		& = (1-P(\tau_s=0))\cdot \frac{1}{1-\pi_1^\prime}\cdot \\
		&\quad (\pi_2^\prime\cdot p_2^2 + \cdots + \pi_k^\prime\cdot p_k^2+ \cdots +\pi_M^\prime\cdot p_M^2).
	\end{split}
	\end{equation*}
	The solution for maximizing $P(\tau_s=1)$ is to make $p_2^2$ be $1$ and the rest to be $0$. Thus, $P(\tau_s=1)\leq \pi_2^\prime$.
	
	Concequently, we can draw the conclusion that 
	\begin{equation*}
	E(\tau_s)=\sum_{t=0}^{\infty} tP(\tau_s=t)\geq \sum_{k=1}^{M} (k-1)\pi_k^\prime
	\end{equation*} 

	Combining the above conclusions, the Bayesian risk considering switching cost in our problem is 
	\begin{equation}
	\label{lower}
	\begin{split}
	{R}(\Gamma)& ={P}_{e}(\Gamma)+cE(\tau|\Gamma)+sE(\tau_s|\Gamma)\\
	& \geq \frac{-c \log c}{I^*(M)}+s\cdot \sum_{k=1}^{M} (k-1)\pi_k^\prime, 
	\end{split}
	\end{equation} 
	where $\pi_k^\prime$ denote the $k^{th}$ highest probability of the priori probability set $\{\pi_m \}_{m=1}^{M}$.
	\end{IEEEproof}
\end{proposition}

\textit{2) Asymptotic Optimality of DBS Policy in Scenario 1} 

In the following part, firstly, Lemma \ref{pe} shows that the error probability $P_e=O(c)$. Secondly, Lemma \ref{t1} to Leamma \ref{ts} shows that the expected switching time of the DBS policy is upper bounded by a constant. Then, Lemma \ref{tc} shows that the expected detection time under the DBS policy is upper bounded by $-(1+o(1))\frac{\log c}{I^*(M)}$. Finally, We will prove that the asymptotic Bayesian risk of DBS policy approaches the asymptotic lower bound.

\begin{lemma}
	\label{pe} 
	In the asymptotic regime, the error probability of DBS policy is upper bounded by:
	\begin{equation}
	\label{process_error}
	P_e \leq (M-1)c.
	\end{equation} 
	\begin{IEEEproof}
		Let $\alpha_{m,j}= P_{m}(\delta = H_j)$ is the probability of declaring $\delta=H_j$ for all $j\neq m$ when $H_m$ is true. Thus, $\alpha_{m}=\sum_{j \neq  m} \alpha_{m,j}$. Note that accepting $H_j$ implies that $S_j(n)\geq -\log c$ in Case I or $S_i(n)\leq \log c$ for all $i\neq j$ in Case II. The stopping rule of single process is the same as the stopping rule of SPRT \cite{Wald}. According to the conclusion of SPRT\cite{Wald}, we can show that for all $H_j \neq  H_m$,
		\begin{equation}
		\alpha_{m}=\sum_{j \neq  m} \alpha_{m,j}\leqslant (M-1)c.
		\end{equation} 
		
		Hence, (\ref{process_error}) follows.
	\end{IEEEproof}
\end{lemma}

Under Hypothesis $H_m$, ${S}_{m}(n)$ is a random walk with positive expected increment $E_{m}(l_m(n))=D(g||f)> 0$ for $m$, while ${S}_{j}(n)$, is a random walk with negative expected increment $E_{m}(l_j(n))=-D(f||g)< 0$ for $j \neq m$. As a result, in Case I, the process $m$ is observed with higher probability than other processes $j$ when $n$ is sufficiently large, and the situation is opposite in Case II. Next, we define a random time $\tau_1^k$. In Case I, $\tau_1^k$ is the last passage time when process $m^{1}(n)$ (the largest observed sum-LLRs process) is always the abnormal process for all $n\geq \tau_{1}^k$ at stage $k$. In Case II, $\tau_1^k$ is the last passage time when process $m^{-1}(n)$ (the smallest observed sum-LLRs process) is always normal for all $n\geq \tau_{1}^k$ at stage $k$. It should be noted that $\tau_1^k$ is not a stopping time and the agent does not know whether $\tau_1^k$ has arrived. In Lemma \ref{t1} below we show that $\tau_1^k$ is sufficiently small with high probability. We will use this result later to establish the upper bound of the actual stopping time $\tau$, the observations time $\tau_c$ and the switching time $\tau_s$ under DBS policy.

\begin{definition}
	\label{def1}
	In Case I, denote $\tau_{1}^k$ as the smallest integer such that $m^{1}(n) \neq j$ for all normal process $j$ and all $n\geq \tau_{1}^k$ in stage $k$. In Case II, denote $\tau_{1}^k$ as the smallest integer such that $m^{-1}(n) \neq m$ for the abnormal process $m$ and all $n\geq \tau_{1}^k$ in stage $k$.
\end{definition}

\begin{lemma}
	\label{t1}  
	If the DBS policy has not been terminated, there exist finite constants $D_1 > 0$ and $\mu_1 >0$ such that
	\begin{equation*}
	{P_m}(\tau_{1}^{k}>n)\leq D_1 e^{-\mu_1 n}
	\end{equation*}
	for all $i=1,2,...,M$ at the corresponding stage.
	\begin{IEEEproof}
		There is only one stage in Case I, our policy observes the $m^{1}(n)$ process and finalizes the stage once ${S}_{m^{1}(n)}(n) \geq -\log c$. Note that 
		\begin{equation}
		\begin{split}
		{P_m}(\tau_{1}^{k}>n) & \leq {P_m}(\max_{j\neq m}\sup_{t\geq n}{S}_{j}(t)-S_m(t)\geq 0)\\
		& \leq \sum_{j\neq m}\sum_{t=n}^{\infty } P_{m}({S}_{j}(t)\geq{S}_{m}(t)).
		\end{split}
		\end{equation}
		
		In each stage of Case II, our policy will choose to observe the $m^{-1}(n)$ process and finalizes the stage once ${S}_{m^{-1}(n)}(n) \leqslant \log c$. Note that 
		\begin{equation}
		\begin{split}
		\label{proof-caseII-1}
		{P_m}(\tau_{1}^{k}>n) & \leq {P_m}(\min_{j\neq m}\inf_{t\geq n}({S}_{m}(t)-{S}_{j}(t))\leq 0)\\
		& \leq \sum_{j\neq m}\sum_{t=n}^{\infty }P_{m}({S}_{j}(t)\geq{S}_{m}(t)).
		\end{split}
		\end{equation}
		
	Therefore, it suffices to show that there exist finite constants $D_1^\prime > 0$ and $\mu_1^\prime >0$ such that $P_{m}({S}_{j}(t)\geq{S}_{m}(t)) \leq D_1^\prime e^{-\mu_1^\prime n}$ for both Case I and Case II. The \cite[Lemma 7]{cohen_bayes} has proved the conclusion for Case I, because the preconditions of DBS policy is the same as the DGF policy\cite{cohen_bayes} in Case I, the conclusion is still valid in this problem. Then, we focus on Case II and show that there exit $D > 0$ and $\mu >0$ such that $P_{m}({S}_{j}(t)\geq{S}_{m}(t)) \leq De^{-\mu n}$. 
	
	Let $N_j(n)\triangleq \sum_{t=1}^n \textbf{1}_{j}(n)$ be the number of times that process $j$ has been observed up to time $n$. Each term in the summation on the RHS of (\ref{proof-caseII-1}) can be upper bounded by 
	\begin{equation}
	\begin{split}
	& P_m(S_j(n)\geq S_m(n) ) \\
	\leq & P_m(S_j(n)\geq S_m(n),N_j(n)\geq \rho n) \\    
	& +  P_m(S_j(n)\geq S_m(n),N_m(n)\geq \rho n) \\
	& + P_m(S_j(n)\geq S_m(n),N_j(n)<\rho n, N_m(n) < \rho n), \\
	\end{split}
	\end{equation}
	where $\rho=\frac{1}{16(M-2)}$ and $0< \rho \leq 1/16$. Specifically, applying (\cite{cohen_bayes}, Lemma 7, (67)) to our model, we can draw the conclusion that there exist finite constants $\gamma_1>0$ and $C_1>0$ such that
	
	\begin{equation}
	\begin{split}
	\label{proof-caseII-2}
	& P_m(S_j(n)\geq S_m(n) ) \\
	\leq & 2C_1e^{-\gamma_1 n} + P_m(S_j(n)\geq S_m(n),N_j(n)<\rho n, N_m(n) < \rho n)\\
	\leq & 2C_1e^{-\gamma_1 n} + \\ & \sum_{r\neq j,m}P_m \left( \tilde{N}_r(n)> \frac{n(1-2\rho)}{M-2},N_j(n)<\rho n, N_m(n) < \rho n \right),
	\end{split}
	\end{equation}
	From the second last inequality to the last inequality, what needs note is that the event $(N_j(n)<\rho n, N_m(n) < \rho n)$ implies that processes $j,m$ are not observed at least $\tilde{n} = n- N_j(n)- N_m(n) \geq n(1-2\rho)$ times. Let $\tilde{N}_r(n)$ be number of times when process $r\neq j,m$ has been probed and process $j,m$ have not been probed up to time $n$. We refer to each such time as $r_{\neq j,m}$-probing time. there exist a process $r\neq j,m$ such that $\tilde{N}_r(n)\geq \frac{\tilde{n}}{M-2}\geq \frac{n(1-2\rho)}{M-2}$. 
	
	It remains to show that the second term in the summation on the RHS of (\ref{proof-caseII-2}) decreases exponentially with $n$.  Due to the difference in the selection rules of DBS policy and DGF policy in Case II, it should be noted that at every $r_{\neq j,m}$-probing time, $S_j(n)\geq S_r(n)$ and $S_m(n)\geq S_r(n)$ must occur in Case II of DBS policy because the agent always observes the process $m^{-1}(n)$. Let $\tilde{t}_1^r, \tilde{t}_2^r,\cdots,\tilde{t}_{\tilde{N}_r(n)}^r$ be the $r\neq j,m$-probing time indices and let $\zeta \triangleq \frac{1-2\rho}{2(M-2)}$. The event $\tilde{N}_r(n) \geq \frac{n(1-2\rho)}{M-2}$ implies that at time $\tilde{t}_{\zeta n}^r$ the inequalities $S_j(\tilde{t}_{\zeta n}^r)\geq S_r(\tilde{t}_{\zeta n}^r)$ and $S_m(\tilde{t}_{\zeta n}^r)\geq S_r(\tilde{t}_{\zeta n}^r)$ must occur. Since $N_r(n)$ is the total number of observations taken from process $r$ up to time $t$, then $N_r(\tilde{t}_{\zeta n}^r)\geq \tilde{N}_r (\tilde{t}_{\zeta n}^r) = \zeta n$. Hence, each term in the summation on the RHS of (\ref{proof-caseII-2}) can be upper bounded by 
	\begin{equation}
	\begin{split}
	\label{proof-caseII-3}
	& P_m\left( N_r(n)> \frac{n(1-2\rho)}{M-2},N_j(n)<\rho n, N_m(n) < \rho n \right)\\
	\leq & P_m \left( \tilde{N}_r(n)>\zeta n,N_j(n)<\rho n \right)+ \\ & P_m\left(\tilde{N}_r(n)>\zeta n,N_m(n)<\rho n \right)\\
	\leq & \sum_{N_r'=\zeta n}^{n} P_m\left( S_j(\tilde{t}_{\zeta n}^r)\geq S_r(\tilde{t}_{\zeta n}^r),N_j(n)<\rho n \right) + \\ & \sum_{N_r'=\zeta n}^{n} P_m\left( S_m(\tilde{t}_{\zeta n}^r)\geq S_r(\tilde{t}_{\zeta n}^r),N_m(n)<\rho n \right). \\
	\end{split}
	\end{equation}

	Using the i.i.d. property of the LLRs across time we have 
	\begin{equation}
	\begin{split}
	\label{step2}
	 & P_m\left( N_r(n)> \frac{n(1-2\rho)}{M-2},N_j(n)<\rho n, N_m(n) < \rho n \right)\\
	\leq & \sum_{N_r'=\zeta n}^{n} \sum_{N_j=0}^{n} P_m \left( \sum_{i=1}^{N_j} l_j(i) \geq \sum_{i=1}^{N_r'}l_r(i) \right) + \\& \sum_{N_r'=\zeta n}^{n} \sum_{N_m=0}^{n} P_m \left( \sum_{i=1}^{N_m} l_m(i) \geq \sum_{i=1}^{N_r'}l_r(i) \right) \\
	\leq & \sum_{q=0}^{n-\zeta n} \sum_{N_j=0}^{\rho n} P_m \left( \sum_{i=1}^{N_j} l_j(i) \geq \sum_{i=1}^{\zeta n+q}l_r(i) \right) + 
	\\ &\sum_{q=0}^{n-\xi n} \sum_{N_m=0}^{\rho n} P_m \left( \sum_{i=1}^{N_m} l_m(i) \geq \sum_{i=1}^{\xi n+q}l_r(i) \right).
	\end{split}
	\end{equation}
	
	Then we need to bound each term in the summation on the RHS of (\ref{step2}). Note that 
	\begin{equation}
	\sum_{i=1}^{\zeta n + q} l_r(i) + \sum_{i=1}^{N_j}-l_j(i) = \sum_{i=1}^{\zeta n + q} \tilde{l}_r(i) + \sum_{i=1}^{N_j}-\tilde{l}_j(i) -D(f||g)(\zeta n+q -N_j),
	\end{equation}
	where
	\begin{equation*}
	\label{llr1}
	\tilde{l}_k(i) = \begin{cases} 
	l_k(i) - D(g||f),  \text{if } k=m,   \\ 
	l_k(i) + D(f||g),  \text{if } k\neq m.   \\     \end{cases}
	\end{equation*}

	Then,
	\begin{equation}
	\begin{split}
	& P_m\left( \sum_{i=1}^{\zeta n +q} l_r(i) \leq \sum_{i=1}^{N_j} l_j(i) \right) \\
	\leq & P_m \left(\sum_{i=1}^{\xi n+q} \tilde{l}_r(i) + \sum_{i=1}^{n'} - \tilde{l}_j(i) \geq D(f||g) (\zeta n +q-N_j) \right)\\
	\leq & P_m \left(\sum_{i=1}^{\xi n+q} \tilde{l}_r(i) + \sum_{i=1}^{n'} - \tilde{l}_j(i) \geq C_1 (\zeta n +q) \right).\\
	\end{split}
	\end{equation}
	
	Applying the Chernoff inequality and using the i.i.d. property of $\tilde{l}_r(i), \tilde{l}_j(i)$ across time, 
	\begin{equation}
	\begin{split}
	& P_m \left(\sum_{i=1}^{\xi n+q} \tilde{l}_r(i) + \sum_{i=1}^{n'} - \tilde{l}_j(i) \geq C_1 (\zeta n +q) \right)\\
	\leq & E_m\left( e^{s\cdot \left(\sum_{i=1}^{\xi n+q} \tilde{l}_r(i) + \sum_{i=1}^{n'} - \tilde{l}_j(i)\right) } \right)\cdot e^{-sC_1(\zeta n +q)} \\
	\leq &  \left( E_m( e^{s\cdot \tilde{l}_r(1)} ) \right)^{\xi n +q} \cdot  \left( E_m( e^{-s\cdot \tilde{l}_j(1)} ) \right)^{N_j}\cdot e^{-sC_1(\zeta n +q)}  \\
	\leq &  \left( E_m( e^{s\cdot( \tilde{l}_r(1)+C_1 ) } ) \right)^{\xi n +q} \cdot  \left( E_m( e^{ s\cdot (-\tilde{l}_j(1)+C_1) } ) \right)^{N_j} \\ & \cdot e^{-s C_1(\zeta n +q)} \cdot e^{-s C_1(\zeta n +q+N_j)}	\\
	\end{split}
	\end{equation}
	for all $s<0$ and $C_1=D(f||g)$. Since $E_m(  \tilde{l}_r(1)+C_1) = C_1>0$ and $E_m(  -\tilde{l}_j(1)+C_1 ) = C_1 >0$ are strictly positive, differentiating the MGFs (Moment Generating Functions) of $l_j(i), l_m(i)$ with respect to $s$ yields strictly positive derivatives at $s=0$. As a result, there exist constants $s<0$ and $\gamma_2>0$ such that $E_m( e^{s\cdot( \tilde{l}_r(1)+C_1 ) } )$ , $ E_m( e^{ s\cdot (-\tilde{l}_j(1)+C_1) } )$ and $e^{sC_1}$ are strictly less than $e^{-\gamma_2}<1$. Hence,
	
	\begin{equation}
	\begin{split}
	& P_m\left( \sum_{i=1}^{\zeta n +q} l_r(i) \leq \sum_{i=1}^{N_j} l_j(i) \right) \\
	\leq &  \left( E_m( e^{s\cdot( \tilde{l}_r(1)+C_1 ) } ) \right)^{\xi n +q} \cdot  \left( E_m( e^{ s\cdot (-\tilde{l}_j(1)+C_1) } ) \right)^{N_j} \\ & \cdot e^{-s C_1(\zeta n +q)} \cdot e^{-s C_1(\zeta n +q+N_j)}	\\
	\leq & e^{-\gamma_2(\zeta n+q)}\cdot e^{-\gamma_2 N_j} \cdot e^{-\gamma_2(\zeta n+q)} \cdot e^{-\gamma_2 (\zeta n + q +N_j)} \\ &= e^{-\gamma_2(3\zeta n + 3q+2N_j)}.
	\end{split}
	\end{equation}
	
	Therefore,
	\begin{equation}
	\begin{split}
	& \sum_{q=0}^{n-\zeta n} \sum_{N_j=0}^{\rho n} P_m \left( \sum_{i=1}^{N_j} l_j(i) \geq \sum_{i=1}^{\zeta n+q}l_r(i) \right) \\
	\leq & \sum_{q=0}^{n-\zeta n} \sum_{N_j=0}^{\rho n}e^{-\gamma_2(3\zeta n + 3q+2N_j)}\\
	\leq & e^{-\gamma_2 \cdot 3\zeta n} \cdot \frac{1}{1-e^{-3\gamma_2}} \cdot \frac{1}{1-e^{-2\gamma_2}} = C_3 e^{-\gamma_3 n},
	\end{split}
	\end{equation}
	where $C_3=\frac{1}{(1-e^{-3\gamma_2)}(1-e^{-2\gamma_2})}$ and $\gamma_3=3\zeta \gamma_2$.
	
	Following the minor modifications, we can prove that there exist finite constants $C_4>0$ and $\gamma_4>0$ such that
	\begin{equation}
	\sum_{q=0}^{n-\xi n} \sum_{N_m=0}^{\rho n} P_m \left( \sum_{i=1}^{N_m} l_m(i) \geq \sum_{i=1}^{\xi n+q}l_r(i) \right) \leq C_4 e^{-\gamma_4 n}.
	\end{equation}
	
	Thus, we complete the proof.
	\end{IEEEproof}
\end{lemma}  

In what follows we define the second random time $\tau_2^k$, where $\tau_2^k$ can be viewed as the last passage time when the switching no longer occurs at stage $k$. It should be noted that $\tau_2^k \geq \tau_1^k$ and the agent does not know whether $\tau_2^k$ has arrived. In Lemma \ref{t2} we show that the total time between $\tau_1^k$ and $\tau_2^k$ is sufficiently small with high probability, which means that the number of switching is limited.

\begin{definition}
\label{def2}
	In Case I, denote $\tau_{2}^k$ as the smallest integer such that the value of $m^{1}(n)$ no longer changes for all $n\geq \tau_{2}^k$ in stage $k$. In Case II, denote $\tau_{2}^k$ as the smallest integer such that the value of $m^{-1}(n)$ no longer changes for all $n\geq \tau_{2}^k$ in stage $k$.
\end{definition}

\begin{definition}
	$n_2^k \triangleq \tau_{2}^k-\tau_{1}^k$ denotes the total amount of time between $\tau_{1}^k$ and $\tau_{2}^k$ in stage $k$.
\end{definition}

\begin{lemma}
	\label{t2}  
	If DBS policy has not been terminated, there exist $D_2 > 0$ and $\mu_2 >0$ such that
	\begin{equation*}
	P_m(n_{2}^{k}>n)\leq D_2e^{-\mu_2 n}
	\end{equation*}
	for all $m=1,2,...,M$ at the corresponding stage.
	\begin{IEEEproof}
		In Case I, we have only one anomalous process. It is obvious that $\tau_{2}^k=\tau_{1}^k$ since there is only one stage, so the conclusion of Lemma \ref{t1} also proves that there exist constants $D_2 > 0$ and $\mu_2 >0$ such that $\textbf{P}_m(n_{2}^{k}>n)\leq D_2e^{-\mu_2 n}$.
		
		In Case II, it can be verified that the $M-1$ normal processes are observed with higher probability than anomalous process by Lemma \ref{t1}. ${S}_{m^{-1}(n)}(n)$ is likely to be a random variable with negative expected increment $E(l_{m^{-1}(n)}(n))< 0$. It is obvious that $S_{m^{-1}(n)}(n)\leq S_{m^{-2}(n)}(n)$, so  $E({S}_{m^{-2}(t)}(t)-{S}_{m^{-1}(t)}(t)) > 0$.
		
		Note that, 
		\begin{equation}
		\begin{split}
		& P_m(n_{2}^{k}>n)\\
		 \leq & P_m(\sup_{t\geq n}l_{m^{-1}(t)}(t+1)+{S}_{m^{-1}(t)}(t)-{S}_{m^{-2}(t)}(t) \geq 0)\\
		\leq &\sum_{t=n}^{\infty }P_m\left( \tilde{l}_{m^{-1}(t)}(t+1)+ {S}_{m^{-1}(t)}(t)-{S}_{m^{-2}(t)}(t) \geq D(f||g) \right)\\
		 \leq &\sum_{t=n}^{\infty } E_m\left(e^{s\cdot \left(\tilde{l}_{m^{-1}(t)}(t+1)+ {S}_{m^{-1}(t)}(t)-{S}_{m^{-2}(t)}(t) - \epsilon \right) } \right) \\ 
		 & \cdot e^{-s(D(f||g)-\epsilon ) } \\
 		\end{split}
		\end{equation}
		for all $s>0$ and $0<\epsilon<D(f||g)$.
		
		The last equality follows by applying the Chernoff bound for each term in the summation on the RHS of the equality. Since $E_m \left( (\tilde{l}_{m^{-1}(t)}(t+1)+ {S}_{m^{-1}(t)}(t)-{S}_{m^{-2}(t)}(t)- \epsilon \right) \leq - \epsilon < 0$ are strictly negative, differentiating the MGFs of $\tilde{l}_{m^{-1}(t)}(t+1)+ {S}_{m^{-1}(t)}(t)-{S}_{m^{-2}(t)}(t)- \epsilon$ with respect to $s$ yields strictly negative derivatives at $s=0$. Hence, there exist $s>0$ and $\gamma>0$ such that $E_m \left( (\tilde{l}_{m^{-1}(t)}(t+1)+ {S}_{m^{-1}(t)}(t)-{S}_{m^{-2}(t)}(t)- \epsilon \right)$ and $e^{-s\epsilon}$ are strictly less than $e^{-\gamma} <1$. Furthermore, there exist $\gamma'>0$ such that each term in the summation on the RHS less or equal to $e^{-\gamma' t}$. Thus, there exist $C_5>0$ and $\gamma_5>0$ such that
		
		\begin{equation}
		\begin{split}
		& \sum_{t=n}^{\infty } E_m\left(e^{s\cdot \left(\tilde{l}_{m^{-1}(t)}(t+1)+ {S}_{m^{-1}(t)}(t)-{S}_{m^{-2}(t)}(t) - \epsilon \right) } \right)\\ 
		&\quad \cdot e^{-s(D(f||g)-\epsilon ) } \\
		\leq & \sum_{t=n}^{\infty } e^{-\gamma' t} \leq C_5 e^{-\gamma_5 n}, 
		\end{split}
		\end{equation}
		which completes the proof.
	\end{IEEEproof}
\end{lemma}

According to the definition \ref{def1} and definition \ref{def2}, we can know that the switchings between two different processes only occur during $\tau_1^k$ and $\tau_2^k$. Based on this, Lemma \ref{switchings} will show that the time of switchings decreases exponentially with $n$.

\begin{lemma}
	\label{switchings}
	Assume that the DBS policy is implemented indefinitely. Then, there exist finite constants $D_3>0$ and $\gamma_3$ such that
	\begin{equation*}
	P_m(\tau_s^k>n) \leq D_3 e^{-\gamma_3 n}
	\end{equation*} 
	for all $m=1,2,...,M$ at the corresponding stage.
	
	\begin{IEEEproof}
		It should be noted that the switching only occurs during $\tau_1^k$ and $n_2^k$, therefore the times of switchings at each stage follow that $\tau_s^k\leq \tau_1^k + n_2^k$. Thus, there exist constants $D_3>0$ and $\gamma_3>0$ such that,
		\begin{equation}
		  \begin{split}
		  P_m(\tau_s^k>n) & \leq P_m(\tau_1^k+n_2^k>n) \\
		  & \leq P_m(\tau_1^k>\frac{n}{2}) + P_m(n_2^k>\frac{n}{2}) \\
		  & \leq D_1 e^{-\gamma_1 n} + D_2 e^{-\gamma_2 n} \\
		  & \leq D_3 e^{-\gamma_3 n},
		  \end{split}
		\end{equation}
		which completes the proof.
	\end{IEEEproof}

\end{lemma}

Below, we difine a random time $\tau_3^k$. In Case I, $\tau_3^k$ is the time when sufficient information for declaring process $m^{1}(n)$ as abnormal process has been gathered. In Case II, denote $\tau_3^k$ as the time when sufficient information for declaring process $m^{-1}(n)$ as normal process has been gathered.

\begin{definition}
	In Case I, denote $\tau_3^k$ as the smallest integer such that $S_{m^1(\tau_3^k)}(\tau_3^k)\geq -\log c$. In Case II, denote $\tau_3^k$ as the smallest integer such that $S_{m^{-1}(\tau_3^k)}(\tau_3^k)\leq \log c$.
\end{definition}

\begin{definition}
	$n_3^k \triangleq \tau_{3}^k-\tau_{2}^k$ denotes the total amount of time between $\tau_{3}^k$ and $\tau_{2}^k$ in stage $k$.
\end{definition}

Combine the above definitions and conclusions, the DBS policy does not occurs switchings during $\tau_2^k$ to $\tau_3^k$. For each process, we will stop observing once $|S_m(n)|\geq -\log c$, it is equivalent to doing SPRT for each process during $n_3^k$. Thus, according to the conclusion of SPRT, we can draw the conclusion that, 

\begin{equation}
\begin{split}
E_m(n_3^k) \leq \begin{cases}
\frac{-\log c}{D(g||f)}, \text{ if Case I}, \\
\frac{-\log c}{D(f||g)}, \text{ if Case II}.
\end{cases}
\end{split}
\end{equation}

Next, we will analyze the expected times of switchings and observations of DBS policy in the asymptotic regime. 

\begin{lemma}
	\label{ts}
	Assume that the DBS policy is implemented indefinitely. Then, there exists $T_s\in (0,\infty)$ such that the expected switching time of the DBS policy $E_m(\tau_s|\text{DBS})$ satisfies
	\begin{equation}
	E_m(\tau_s|\text{DBS})\leq T_s.
	\end{equation}
	
	\begin{IEEEproof}
	We first focus on one stage in the entire detection procedure. Then,
	\begin{equation}
	\begin{split}
	E_m(\tau_s^k) & = \sum_{t=1}^{\infty} t\cdot P_m(\tau_s^k=t) \\
	& \leq \sum_{t=1}^{\infty} t\cdot P_m(\tau_s^k>t-1)\\
	& \leq \sum_{t=1}^{\infty} t \cdot D_3 e^{-\gamma_3(t-1)} \\
	& = D_3 e^{\gamma_3} \sum_{t=1}^{\infty} te^{-\gamma_3 t} = D_3 (1-e^{-\gamma_3})^2,
	\end{split}
	\end{equation}	
	where $0<D_3<\infty$ and $\gamma_3>0$. Thus, there exist $T_s\in (0,\infty)$ such that
	\begin{equation}
	\begin{split}
	E_m(\tau_s)&=E_m( \sum_k \tau_s^k ) = \sum_k E_m(\tau_s^k) \\
	& \leq \sum_k D_3 (1-e^{-\gamma_3})^2 \leq T_s,
	\end{split}
	\end{equation}
	which completes the proof of Lemma \ref{ts}.
	\end{IEEEproof}
\end{lemma}

It remains to show that the expected number of observations of the DBS policy. 

\begin{lemma}
	\label{tc}  
	In the asymptotic regime, the expected detection time $\tau_c$ under the DBS policy is bounded by:
	\begin{equation}
	E(\tau_c) \leq -(1+o(1))\frac{\log c}{I^*(M)} \text{ as } c\rightarrow 0.
	\end{equation}
	
	\begin{IEEEproof}
		Following the above definitions, in Case I, 
		\begin{equation}
		\begin{split}
		E_m(\tau_c)  & =E_m(\tau _3^1)\\
		& = E_m(\tau_1^1 + n_2^1) + E_m(n_3^1) \\
		& \leq T_s + \frac{-\log c}{D(g||f)} \\
		& \leq -(1+o(1))\frac{\log c}{D(g||f)} \text{ as } c \rightarrow 0.
		\end{split}
		\end{equation}
		
		In Case II,
		\begin{equation}
		\begin{split}
		E_m(\tau_c)  & =E_m\left( \sum_{k=1}^{M-1} \tau_3^k \right)\\
		& =\sum_{k=1}^{M-1} \left( E_m(\tau_1^k + n_2^k) + E_m(n_3^k) \right) \\
		& \leq \sum_{k=1}^{M-1} \left( T_s + \frac{-\log c}{D(f||g)} \right)\\
		& \leq -(1+o(1))\frac{(M-1)\log c}{D(f||g)} \text{ as } c \rightarrow 0.
		\end{split}
		\end{equation}
		Combine the definition of (\ref{imk}), we complete the proof.
		\end{IEEEproof}
\end{lemma}

\begin{lemma}
	\label{dbs-risk}
	In the asymptotic regime, the Bayesian risk under the DBS policy is upper bounded by
	\begin{equation}
	R_m(\text{DBS})\leq O(c)+(1+o(1))\cdot\frac{-c\log c}{I^*(M)}+s\cdot T_s
	\end{equation}
	for all $m=1,2,...,M$.
	
	\begin{IEEEproof}
		Combining Lemmas \ref{ts}, \ref{tc} completes the proof.
	\end{IEEEproof}
\end{lemma}

Combining Proposition 1 and Lemma \ref{dbs-risk} yields the asymptotic optimality of DBS policy, presented in Theorem 1,
\begin{equation}
\label{asymptotic}
\begin{split}
\lim_{c\rightarrow 0} \frac{R(\text{DBS})}{\inf_\Gamma R(\Gamma)} & \leq \lim_{c\rightarrow 0} \frac{c \cdot \frac{-\log c}{I^*(M)} + s\cdot T_s}{c \cdot \frac{-\log c}{I^*(M)} + s\cdot \sum_{k=1}^{M}(k-1)\pi_k } \\ 
& = \lim_{c\rightarrow 0} 1+ \frac{s\cdot (T_s -\frac{M-1}{2} ) }{c \cdot \frac{-\log c}{I^*(M)} + s\cdot \sum_{k=1}^{M}(k-1)\pi_k }\\ 
& = \lim_{c\rightarrow 0} 1+ \frac{T_s -\frac{M-1}{2}  }{\frac{-c \log c}{s} \cdot \frac{1}{I^*(M)} +\sum_{k=1}^{M}(k-1)\pi_k }.\\ 
\end{split}
\end{equation}
Note that $s=o(-c\log c)$ as conditioned by Theorem 1, we can know that $\lim_{c\rightarrow 0} \frac{-c\log c}{s}\rightarrow \infty$. Then, 
\begin{equation}
\begin{split}
\lim_{c\rightarrow 0} \frac{R(\text{DBS})}{\inf_\Gamma R(\Gamma)} & = \lim_{c\rightarrow 0} 1+ \frac{T_s -\frac{M-1}{2}  }{\frac{-c \log c}{s} \cdot \frac{1}{I^*(M)} +\sum_{k=1}^{M}(k-1)\pi_k }\\ 
& = 1+ \frac{T_s -\frac{M-1}{2}  }{\infty  \cdot \frac{1}{I^*(M)} +  \sum_{k=1}^{M}(k-1)\pi_k } = 1,
\end{split}
\end{equation}
which completes the proof. 

\subsection{The Proof of Theorem 2}
		
	The proof follows a similar line of arguments as in the proof of Theorem 1. Since the ratio of $s/-c \log c$ does not affect the expected number of observations and switchings, the Lemma \ref{tc} and \ref{ts} still holds for Scenario 2 of the DBS policy. Thus,
	\begin{equation}
	\begin{split}
	\lim_{c\rightarrow 0} \frac{R(\text{DBS})}{\inf_\Gamma R(\Gamma)} & \leq \lim_{c\rightarrow 0} \frac{c \cdot \frac{-\log c}{I^*(M)} + s\cdot T_s}{c \cdot \frac{-\log c}{I^*(M)} + s\cdot \sum_{k=1}^{M}(k-1)\pi_k }
	\\ & = \lim_{c\rightarrow 0} 1+ \frac{s\cdot (T_s -\sum_{k=1}^{M}(k-1)\pi_k ) }{c \cdot \frac{-\log c}{I^*(M)} + s\cdot \sum_{k=1}^{M}(k-1)\pi_k }
	\\ & = \lim_{c\rightarrow 0} 1+ \frac{T_s -\sum_{k=1}^{M}(k-1)\pi_k  }{\frac{-c \log c}{s} \cdot \frac{1}{I^*(M)} +\sum_{k=1}^{M}(k-1)\pi_k}.
	\end{split}
	\end{equation}
	Differ from the condition that $s=o(-c\log c)$ as $c\rightarrow 0$ in Scenario 1, we assume that $s=\Omega(-c\log c)$ as $c$ approaches $0$ in Scenario 2, which means that $\lim_{c\rightarrow 0} \frac{-c\log c}{s}\geq 0$. Thus,
	\begin{equation}
	\begin{split}
	\lim_{c\rightarrow 0} \frac{R(\text{DBS})}{\inf_\Gamma R(\Gamma)} & = \lim_{c\rightarrow 0} 1+ \frac{T_s -\sum_{k=1}^{M}(k-1)\pi_k  }{\frac{-c \log c}{s} \cdot \frac{1}{I^*(M)} +\sum_{k=1}^{M}(k-1)\pi_k}
	\\ & \leq  1+ \frac{T_s -\sum_{k=1}^{M}(k-1)\pi_k  }{ \sum_{k=1}^{M}(k-1)\pi_k }
	\\ & \leq \frac{T_s}{\sum_{k=1}^{M}(k-1)\pi_k} < \infty,
	\end{split}
	\end{equation} 
	which completes the proof.

\ifCLASSOPTIONcaptionsoff
\newpage
\fi

\bibliographystyle{IEEEtran}
\bibliography{refs}

\begin{thebibliography}{10}
\providecommand{\url}[1]{#1}
\csname url@samestyle\endcsname
\providecommand{\newblock}{\relax}
\providecommand{\bibinfo}[2]{#2}
\providecommand{\BIBentrySTDinterwordspacing}{\spaceskip=0pt\relax}
\providecommand{\BIBentryALTinterwordstretchfactor}{4}
\providecommand{\BIBentryALTinterwordspacing}{\spaceskip=\fontdimen2\font plus
\BIBentryALTinterwordstretchfactor\fontdimen3\font minus
  \fontdimen4\font\relax}
\providecommand{\BIBforeignlanguage}[2]{{%
\expandafter\ifx\csname l@#1\endcsname\relax
\typeout{** WARNING: IEEEtran.bst: No hyphenation pattern has been}%
\typeout{** loaded for the language `#1'. Using the pattern for}%
\typeout{** the default language instead.}%
\else
\language=\csname l@#1\endcsname
\fi
#2}}
\providecommand{\BIBdecl}{\relax}
\BIBdecl

\bibitem{ChenDa}
D.~Chen, Q.~Huang, H.~Feng, Q.~Zhao, and B.~Hu, ``Active anomaly detection with
  switching cost,'' in \emph{ICASSP 2019-2019 IEEE International Conference on
  Acoustics, Speech and Signal Processing (ICASSP)}.\hskip 1em plus 0.5em minus
  0.4em\relax IEEE, 2019, pp. 5346--5350.

\bibitem{cd_1}
H.~Chernoff, ``Sequential design of experiments,'' \emph{The Annals of
  Mathematical Statistics}, vol.~30, no.~3, pp. 755--770, 1959.

\bibitem{Wald}
A.~Wald, ``Sequential analysis.'' 1947.

\bibitem{Bessler}
S.~A. Bessler, ``Theory and applications of the sequential design of
  experiments, k-actions and infinitely many experiments. part i. theory,''
  Stanford Univ CA Applied Mathematics and Statistics Labs, Tech. Rep., 1960.

\bibitem{Naghshvar_2010}
M.~Naghshvar and T.~Javidi, ``Active m-ary sequential hypothesis testing,'' in
  \emph{2010 IEEE International Symposium on Information Theory}.\hskip 1em
  plus 0.5em minus 0.4em\relax IEEE, 2010, pp. 1623--1627.

\bibitem{Naghshvar_2013}
M.~Naghshvar, T.~Javidi \emph{et~al.}, ``Active sequential hypothesis
  testing,'' \emph{The Annals of Statistics}, vol.~41, no.~6, pp. 2703--2738,
  2013.

\bibitem{Naghshvar-information}
M.~Naghshvar and T.~Javidi, ``Information utility in active sequential
  hypothesis testing,'' in \emph{2010 48th Annual Allerton Conference on
  Communication, Control, and Computing (Allerton)}.\hskip 1em plus 0.5em minus
  0.4em\relax IEEE, 2010, pp. 123--129.

\bibitem{Naghshvar-performance}
M.~Naghshvar and T.~Javidi, ``Performance bounds for active sequential
  hypothesis testing,'' in \emph{2011 IEEE International Symposium on
  Information Theory Proceedings}.\hskip 1em plus 0.5em minus 0.4em\relax IEEE,
  2011, pp. 2666--2670.

\bibitem{Naghshvar-sequential}
M.~Naghshvar and T.~Javidi, ``Sequentiality and adaptivity gains in active
  hypothesis testing,'' \emph{IEEE Journal of Selected Topics in Signal
  Processing}, vol.~7, no.~5, pp. 768--782, 2013.

\bibitem{Nitinawarat-2012}
S.~Nitinawarat, G.~K. Atia, and V.~V. Veeravalli, ``Controlled sensing for
  hypothesis testing,'' in \emph{2012 IEEE International Conference on
  Acoustics, Speech and Signal Processing (ICASSP)}.\hskip 1em plus 0.5em minus
  0.4em\relax IEEE, 2012, pp. 5277--5280.

\bibitem{Nitinawarat-2013}
S.~Nitinawarat, G.~K. Atia, and V.~V. Veeravalli, ``Controlled sensing for
  multihypothesis testing,'' \emph{IEEE Transactions on Automatic Control},
  vol.~58, no.~10, pp. 2451--2464, 2013.

\bibitem{Nitinawarat-2015}
S.~Nitinawarat and V.~V. Veeravalli, ``Controlled sensing for sequential
  multihypothesis testing with controlled markovian observations and
  non-uniform control cost,'' \emph{Sequential Analysis}, vol.~34, no.~1, pp.
  1--24, 2015.

\bibitem{cohen_bayes}
K.~Cohen and Q.~Zhao, ``Active hypothesis testing for anomaly detection,''
  \emph{IEEE Transactions on Information Theory}, vol.~61, no.~3, pp.
  1432--1450, 2015.

\bibitem{cohen_cost1}
K.~Cohen, Q.~Zhao, and A.~Swami, ``Optimal index policies for anomaly
  localization in resource-constrained cyber systems,'' \emph{IEEE Transactions
  on Signal Processing}, vol.~62, no.~16, pp. 4224--4236, 2014.

\bibitem{cohen_tree}
C.~Wang, K.~Cohen, and Q.~Zhao, ``Active hypothesis testing on a tree: Anomaly
  detection under hierarchical observations,'' in \emph{International Symposium
  on Information Theory (ISIT)}.\hskip 1em plus 0.5em minus 0.4em\relax IEEE,
  2017, pp. 993--997.

\bibitem{cohen_cost2}
K.~Cohen and Q.~Zhao, ``Asymptotically optimal anomaly detection via sequential
  testing,'' \emph{IEEE Transactions on Signal Processing}, vol.~63, no.~11,
  pp. 2929--2941, 2015.

\bibitem{cohen_hete}
B.~Huang, K.~Cohen, and Q.~Zhao, ``Active anomaly detection in heterogeneous
  processes,'' \emph{IEEE Transactions on Information Theory}, vol.~65, no.~4,
  pp. 2284--2301, 2018.

\bibitem{cohen_costnonline}
A.~Gurevich, K.~Cohen, and Q.~Zhao, ``Sequential anomaly detection under a
  nonlinear system cost,'' \emph{IEEE Transactions on Signal Processing},
  vol.~67, no.~14, pp. 3689--3703, 2019.

\bibitem{AHT2020-KOBI}
B.~Hemo, T.~Gafni, K.~Cohen, and Q.~Zhao, ``Searching for anomalies over
  composite hypotheses,'' \emph{IEEE Transactions on Signal Processing},
  vol.~68, pp. 1181--1196, 2020.

\bibitem{AHT2014}
A.~Tartakovsky, I.~Nikiforov, and M.~Basseville, \emph{Sequential analysis:
  Hypothesis testing and changepoint detection}.\hskip 1em plus 0.5em minus
  0.4em\relax Chapman and Hall/CRC, 2014.

\bibitem{AHT2017-1}
F.~Cecchi and N.~Hegde, ``Adaptive active hypothesis testing under limited
  information,'' in \emph{Advances in Neural Information Processing Systems},
  2017, pp. 4035--4043.

\bibitem{AHT2020-1}
D.~Kartik, A.~Nayyar, and U.~Mitra, ``Testing for anomalies: Active strategies
  and non-asymptotic analysis,'' \emph{arXiv preprint arXiv:2005.07696}, 2020.

\bibitem{AHT2020-2}
M.-C. Chang and M.~R. Bloch, ``Evasive active hypothesis testing,'' in
  \emph{2020 IEEE International Symposium on Information Theory (ISIT)}.\hskip
  1em plus 0.5em minus 0.4em\relax IEEE, 2020, pp. 1248--1253.

\bibitem{Sluggish}
N.~K. Vaidhiyan and R.~Sundaresan, ``Active search with a cost for switching
  actions,'' in \emph{2015 Information Theory and Applications Workshop
  (ITA)}.\hskip 1em plus 0.5em minus 0.4em\relax IEEE, 2015, pp. 17--24.

\end{thebibliography}
\end{document}